\documentclass[aps,prd,11pt,nofootinbib,onecolumn,eqsecnum,showpacs,showkeys,superscriptaddress,preprintnumbers,altaffilletter,longbibliography]{revtex4-2}
\usepackage{graphicx}
\usepackage{amsmath}
\usepackage{multirow}
\usepackage{hyperref} 
\usepackage{soul}
\usepackage{graphicx}
\usepackage{dcolumn}
\usepackage{amssymb}
\usepackage{amsfonts}
\usepackage{amsbsy}
\usepackage{color}
\usepackage{rotating}
\usepackage[english]{babel}
\usepackage{multirow}
\usepackage{float}
\usepackage{cleveref}
\usepackage{diagbox} 
\usepackage{subcaption}
\usepackage{ulem}
\usepackage{standalone}
\usepackage{import}
\usepackage{tikz} 
\tikzset{
  pics/carc/.style args={#1:#2:#3}{
    code={
      \draw[pic actions] (#1:#3) arc(#1:#2:#3);
    }
  }  
}

\usepackage{booktabs}

\newcolumntype{x}[1]{>{\centering\arraybackslash\hspace{0pt}}p{#1}}

\newcommand{\be}{\begin{equation}}
\newcommand{\ee}{\end{equation}}
\newcommand{\bea}{\begin{eqnarray}}
\newcommand{\eea}{\end{eqnarray}}
\newcommand{\beal}{\begin{aligned}}
\newcommand{\eeal}{\end{aligned}}
\newcommand{\bi}{\begin{itemize}}
\newcommand{\ei}{\end{itemize}}

\hypersetup{
    colorlinks=true,
    linkcolor=blue,
    filecolor=magenta,
    urlcolor=cyan,
    citecolor=cyan
}

\begin{document}


\title{Optical and orbital characterization of spherically symmetric static black holes of self-gravitating new nonlinear electrodynamics model}

\author{{\.I}lim {\.I}rfan {\c C}imdiker}
\email{ilim.cimdiker@phd.usz.edu.pl}
\affiliation{Institute of Physics, University of Szczecin,
Wielkopolska 15, 70-451 Szczecin, Poland}
\affiliation{Astrophysics Research Center, the Open University of Israel, Raanana 4353701, Israel}

\author{Ali {\"O}vg{\"u}n}
\email{ali.ovgun@emu.edu.tr}
\affiliation{Physics Department, Eastern Mediterranean University, Gazimagusa 99628 via Mersin 10, Turkiye}

\author{Yosef Verbin}
\email{verbin@openu.ac.il }
\affiliation{Astrophysics Research Center, the Open University of Israel, Raanana 4353701, Israel}

\begin{abstract}
Horizon scale imaging and precision lensing have turned black holes into quantitative laboratories for strong gravity and for non standard electromagnetic physics. We study the optical appearance and orbital dynamics of a new class of static spherically symmetric black holes sourced by a Palatini inspired nonlinear electrodynamics model, minimally coupled to Einstein-Hilbert gravity. Using a unified geodesic analysis, we identify the key radii that organize the strong field phenomenology. For photons we determine the unstable photon sphere, the associated critical capture threshold, and the resulting shadow size for a distant observer, and we map how these observables respond to the charge and to the nonlinearity index $n$. For massive probes we compute circular orbits and the innermost stable circular orbit, clarifying the departure from the Schwarzschild and Reissner-Nordstr\"om cases. We then connect to classical tests by evaluating the light deflection angle and periastron advance, providing additional diagnostics that complement the shadow. Our results furnish a practical reference model for confronting first order nonlinear electrodynamics black holes with current and forthcoming imaging and lensing data.
\end{abstract}

\date{\today}

\keywords{Black holes; Nonlinear electrodynamics; Light deflection; Shadows; Particle orbits.}

\pacs{95.30.Sf, 04.70.-s, 97.60.Lf, 04.50.Kd }

\maketitle

\section{Introduction}
\label{intro}

Black holes (BHs) provide an unparalleled arena to confront general relativity (GR) with
observations in the strong-field regime. The past decade has witnessed a qualitative shift
from indirect evidence to horizon-scale probes: precision astrometry of stellar orbits around
Sgr~A$^\ast$ has revealed relativistic orbital precession consistent with GR expectations
\cite{Gravity2020S2}, while the Event Horizon Telescope (EHT) has delivered horizon-scale
images of the shadow-casting region of M87$^\ast$ and Sgr~A$^\ast$
\cite{EHTM87I2019,EHTSgrAI2022}. Complementary frameworks have developed the use of horizon-scale imaging and strong-field electromagnetic probes to test the Kerr hypothesis and constrain non-GR deformations across a broad class of scenarios, including parameterized metrics and environment-sensitive shadow models \cite{Johannsen:2010ru,Broderick:2013rlq,Johannsen:2015hib,Bambi:2015kza,Younsi:2016azx,Grenzebach:2014fha,Hennigar:2018hza,Kuang:2022ojj,Bronzwaer:2021lzo,Vagnozzi:2020quf,Chen:2022nbb}. These developments motivate a systematic exploration of
how beyond-GR effects and non-standard matter sectors can imprint themselves on
null geodesics, photon regions, lensing observables, and shadow morphology, thereby enabling
model selection (or exclusion) with increasing observational leverage. Recent work argues that ``spacetime collisions'' can exhibit a distinctly electrodynamic character, motivating strong field tests that treat electromagnetic nonlinearities as dynamical players rather than small corrections \cite{Boyeneni:2025tsx}.

From a theoretical standpoint, BH shadows encode the boundary between captured photon trajectories and those reaching a distant observer. In asymptotically flat
spacetimes this boundary is closely tied to critical impact parameters and the structure of
photon spheres or photon regions \cite{Synge1966,PerlickTsupko:2022,CunhaHerdeiro:2018}.
In parallel, gravitational lensing in both weak and strong deflection regimes provides
complementary observables, including deflection angles, relativistic images, and time delays,
whose analytic characterization has proven especially powerful for discriminating between
nearby spacetimes \cite{Bozza2002StrongField,Virbhadra:1999nm,Claudel:2000yi,Virbhadra:2008ws,Capozziello:2025wwl}. The modern literature emphasizes that,
beyond the idealized ``shadow'' of geometric optics, observable features such as photon rings
and lensing rings depend on astrophysical emission models; nevertheless, the underlying
geometric criticality remains a robust diagnostic of the metric in the near-horizon region
\cite{Gralla:2019xty,PerlickTsupko:2022}. A concrete example in this direction is the rotating Einstein-Euler-Heisenberg black hole, where nonlinear electrodynamics modifies both the shadow morphology and the quasinormal mode spectrum in a correlated way \cite{Lambiase:2024lvo}. Related analyses of nonlinearly charged black holes connect horizon scale imaging with accretion physics by jointly modeling the shadow and thin disk observables across the model parameter space \cite{Uniyal:2023inx}.  

A particularly well-motivated class of departures from the Einstein--Maxwell paradigm arises when the electromagnetic sector is promoted to a nonlinear electrodynamics (NLED) theory. Historically, NLED was introduced to regularize diverging classical fields and self-energies (as in Born--Infeld electrodynamics) and it also appears as an effective description of quantum vacuum polarization in strong fields (Euler--Kockel--Heisenberg), thereby providing a controlled framework for strong-field phenomenology \cite{BornInfeld:1934,Euler-Kockel,HeisenbergEuler:1936,Sorokin:2021tge}. Complementary frameworks have developed the use of horizon-scale imaging and strong-field electromagnetic probes to test the Kerr hypothesis and constrain non-GR deformations across a broad class of scenarios, including parameterized metrics and environment-sensitive shadow models \cite{Johannsen:2010ru,Broderick:2013rlq,Johannsen:2015hib,Bambi:2015kza,Younsi:2016azx,Grenzebach:2014fha,Hennigar:2018hza,Kuang:2022ojj,Bronzwaer:2021lzo,Vagnozzi:2020quf,Chen:2022nbb}. When coupled to gravity, NLED can significantly modify black-hole spacetimes and their causal/optical structure \cite{Novello:2000km,Gibbons:2000xe,Gonzalez:2009nn,Toshmatov:2018tyo}, and in particular it can support regular (non-singular) geometries, as famously demonstrated by the interpretation of the Bardeen spacetime within Einstein--NLED \cite{AyonBeatoGarcia:1998,Ayon-Beato:1998hmi,Ayon-Beato:2000mjt,Ayon-Beato:1999kuh,Bronnikov:2000vy,Dymnikova:2004zc,Balart:2014cga,Fan:2016hvf}. Regular/NLED-supported solutions and their generalizations have also been constructed in broader settings, including higher-dimensional and Lovelock gravities and related nonlinear Maxwell sectors, enlarging the space of analytic strong-field templates beyond four-dimensional Einstein--Hilbert theory \cite{Hassaine:2008pw,Maeda:2008ha,Cataldo:1999wr,Cataldo:2000ns}. Beyond regularity, NLED theories exhibit additional structural features, including electric--magnetic duality rotations and $SL(2,\mathbb{R})$-type invariances in axion--dilaton extensions, which further enrich their gravitational sector \cite{Gibbons:1995cv,Gibbons:1995ap}. Moreover, NLED light propagation  
is governed by an effective optical geometry, so that NLED light rays can deviate from metric null geodesics, inducing characteristic corrections to lensing and shadow observables \cite{Novello:1999pg,DeLorenci:2000yh}. These optical effects have direct phenomenological counterparts in lensing/shadow studies of Born--Infeld-type black holes and plasma-modified propagation, strengthening the link between NLED microphysics and observable ray-tracing signatures \cite{Eiroa:2005ag,Perlick:2015vta}. Moreover, the regularity program has been extended to rotating spacetimes and to alternative constructions where NLED fields source nonsingular rotating black holes, with corresponding implications for photon regions and shadow shapes \cite{Ghosh:2014pba,Toshmatov:2017zpr,Dymnikova:2015hka,Kumar:2020yem,Toshmatov:2021fgm}. These motivations become even sharper in the EHT era: horizon-scale imaging of M87$^\ast$ and Sgr~A$^\ast$ provides direct constraints on non-Einsteinian/charged compact-object scenarios \cite{EHTM87I2019,EventHorizonTelescope:2019ggy,EHTSgrAI2022}, with dedicated analyses showing that the 2017 EHT data can bound effective charges and related couplings \cite{EventHorizonTelescope:2021dqv,Vagnozzi:2023} and that magnetically charged/NLED black holes can be confronted with EHT measurements \cite{Allahyari:2019jqz}. In parallel, analytic developments have clarified the relation between the shadow edge, photon rings, and lensing rings as robust probes of near-horizon geometry \cite{Gralla:2019xty,Cunha:2018acu,PerlickTsupko:2022}, while recent studies emphasize that NLED effects can imprint coherently across several channels such as; shadows, deflection angles, quasinormal spectra, and greybody factors, enabling multi-observable consistency tests \cite{Okyay:2021nnh,Uniyal:2022vdu}. Related model-building directions include NLED coupled to modified-gravity sectors (e.g. Palatini $f(R)$ or $f(R)$-NLED realizations), as well as other regularization mechanisms such as higher-curvature (quadratic) gravity, which collectively broaden the landscape of nonsingular compact objects worth confronting with strong-field observables \cite{Olmo:2011ja,Rodrigues:2015ayd,Rodrigues:2018bdc,Berej:2006cc,Capozziello:2024ucm}.

In this context, recently a new family of nonlinear gauge-field theories has been constructed
by adopting a first-order (Palatini-inspired) formulation for linear and nonlinear electrodynamics, in which the
field strength and potential are treated as \emph{a priori} independent variables {\cite{Rasanen-Verbin2022,Verbin:2025PNLED}. Within this
framework, Verbin {\it et al.} \cite{Verbin:2025PNLED} obtained new self-gravitating solutions, including spherically
symmetric black holes, and identified a``Palatini Inspired NLED'' (PINLED) model as a minimal and analytically
tractable representative exhibiting novel features in both flat space and in the self-gravitating
case. Since the phenomenological discriminant power of such
solutions ultimately hinges on observables, it is timely to develop a detailed optical and
orbital characterization of these geometries.

The aim of this paper is to provide a systematic analysis of the gravitational optics and kinematics of the
static, spherically symmetric  black-hole solutions of a family of PINLED models minimally coupled to
Einstein--Hilbert gravity. Concretely, we (i) determine
the photon sphere structure and the associated critical impact parameter controlling capture
versus escape; (ii) compute the shadow radius as perceived by an observer at infinity and
map its dependence on the NLED charge and the nonlinearity index $n$; and (iii) complement
these horizon-scale diagnostics with classical tests such as light deflection and periastron
precession, enabling cross-validation against both weak field and strong-field constraints.
Our results are presented in a form suited for direct comparison with the Schwarzschild (S) and Reissner-Nordstr\"om (RN) limits, and for integration into broader frameworks that confront
modified-gravity/NLED scenarios with EHT and lensing data \cite{EHTM87I2019,EHTSgrAI2022,Bozza2002StrongField,EventHorizonTelescope:2021dqv,Vagnozzi:2023}.

Beyond furnishing concrete predictions for a new analytic family of NLED black holes, the
broader importance of this work is methodological and phenomenological: it illustrates how
first-order NLED constructions translate into sharp optical signatures, and it strengthens the
bridge between model-building in strong-field gravity and the rapidly maturing observational
program targeting horizon-scale structure. As data quality improves, such theoretically
controlled templates will be essential for turning shadow and lensing measurements into
precision tests of fundamental physics. 

 \section{ PINLED Black Holes}
 \label{pinledblackholes}
 The Palatini inspired  variant of NLED  (PINLED) stands on the basic principle that a theory of vector field may be constructed in terms of a gauge potential $A_\mu$ and an anti-symmetric tensor field $P_{\mu\nu}$ which are independent of each other. The theory is of first order in the sense that the Lagrangian contains only first derivatives of $A_\mu$ through the gauge invariant combination $F_{\mu\nu} \equiv \partial_\mu A_\nu -\partial_\mu A_\nu$ (or $F \equiv dA$). In order to obtain the field equations of the theory, the dynamical variables $A_\mu$ and $P_{\mu\nu}$  are varied independently of each other.

This is actually a family of Lagrangians which are determined by an arbitrary {\it defining function} of the first 4 of the 6  scalars which may be constructed as bilinear products of $F_{\mu\nu}$ and $P_{\mu\nu}$ and their duals $^*F^{\mu\nu}=\epsilon^{\mu\nu\rho\sigma} F_{\rho\sigma}/2\sqrt{-g}\,\,$ and $^*P^{\mu\nu}$ (defined analogously): 
\begin{equation}
Y=P^{\mu\nu}F_{\mu\nu} \; , \;\;\ Z=P^{\mu\nu}P_{\mu\nu} \; , \;\;\  \Upsilon =F_{\mu\nu}\,^{*}P^{\mu\nu} \; , \;\;\ \Omega=P_{\mu\nu}\,^{*}P^{\mu\nu} \; , \;\;\ X=F^{\mu\nu}F_{\mu\nu}\; , \;\;\  \Xi =F_{\mu\nu}\,^{*}F^{\mu\nu}.
\label{FourInvariants}
\end{equation}
 We write therefore the Lagrangian density ($J^\mu$ is a possible current):
\begin{equation}
\mathcal{L}_{_{\rm PINLED}}=  \mathcal{K}(Z,\Omega , Y, \Upsilon)- J^\mu A_\mu\ ,
\label{LagNLED1stOrderGen}
\end{equation}
which gives rise to the following field equations:
\begin{equation}
-2\nabla_\mu \left(\mathcal{K}_Y P^{\mu\nu}+\mathcal{K}_\Upsilon \,^{*}P^{\mu\nu}\right)=J^\nu    \;\;\; , \;\;\;\;\
2 \left(\mathcal{K}_Z P^{\mu\nu}+\mathcal{K}_\Omega \,^{*}P^{\mu\nu}\right)+
\mathcal{K}_Y F^{\mu\nu}+\mathcal{K}_\Upsilon \,^{*}F^{\mu\nu}=0\ .
\label{FEqsNLED1stGen}
\end{equation}
The defining function is of course not entirely arbitrary, but must obey some physical constraints. The conventional practice in the first order formulation of NLED is to impose the condition that the first order theory with the defining function $ \mathcal{K}(Z,\Omega , Y, \Upsilon)$ is equivalent to a certain 2nd order NLED theory \cite{Plebanski1970,Verbin:2025PNLED}. However, the new PINLED family is a wider family of NLED Lagrangians free of this limiting constraint \cite{Rasanen-Verbin2022,Verbin:2025PNLED}. 

As mentioned above, there are still physical constraints to be fulfilled, but we will not deviate here for a full study of this issue. We will specify only one example which is the existence of the Maxwell limit. This will be evidently  the case in the 
 simple PINLED Lagrangian which we choose to study here from now on. In this Lagrangian the nonlinearity originates from the $Y^n$ term of the otherwise linear defining function
$
 \mathcal{K}(Z,\Omega , Y, \Upsilon) = Z/4- Y/2 +\gamma Y^n/(2n)    \ .
$
More explicitly, we write the Lagrangian of the  ``PINLED $Y^n$ model'' (or ``$Y^n$ model'' for short) as
\begin{equation}
\mathcal{L}_{Y^n}^{(1)}=\frac{1}{4}P^{\mu\nu}P_{\mu\nu}-\frac{1}{2}P^{\mu\nu}F_{\mu\nu}+\frac{\gamma}{2n}(P^{\mu\nu}F_{\mu\nu})^n  -J^\mu A_\mu \ ,
\label{LmodelYn}
\end{equation}
where $\gamma$ is a real parameter. It is clear that for $\gamma=0$, this reduces to the 1st order version of Maxwell's theory and $P_{\mu\nu}=F_{\mu\nu}$.

Now, variations with respect to $A_\mu$ and $P_{\mu\nu}$ give the modified source-full ``Maxwell'' equations and the $P$-$F$ relations:
\begin{equation}
\nabla_\mu\left(W(Y)P^{\mu\nu}\right)=J^\nu \; \;\; , \;\;\;\; P_{\mu\nu}= W(Y) F_{\mu\nu} 
\; , \quad \text{where} \quad  W(Y)\equiv 1-\gamma Y^{n-1} \ .
\label{FEqsModelYn}
\end{equation}
 Now we add the Einstein-Hilbert term $R/2\kappa$ to the $Y^n$--Lagrangian for allowing a dynamical gravitational field which solves the Einstein equations
\begin{equation}
G_{\mu\nu}=-\kappa T_{\mu\nu} \;\; , \;\;\;  T_{\mu\nu}=-W^2 F_{\mu\alpha}F_{\nu}^{\ \alpha}-\frac{g_{\mu\nu}}{2}\left[ \frac{n-2}{2n}\,W -  \frac{n-1}{n} \right]Y .
 \label{EinsteinEqsGeneralYn}
\end{equation}
Next we assume  spherically-symmetric electrostatic solutions of the field equations \eqref{FEqsModelYn}--\eqref{EinsteinEqsGeneralYn}. This means that the only non-vanishing field components are $F_{tr}(r)=-F_{rt}(r)$ and $P_{tr}(r)=-P_{rt}(r)$ and  the metric tensor can be written in the form $g_{\mu\nu}= diag(f(r), -1/f(r), -r^2, -r^2 \sin^{2}\theta )$. The NLED field equations can be solved analytically for $F_{tr}$ and $P_{tr}$ in parametric form with the variable $Y$ in the role of the parameter \cite{Verbin:2025PNLED}:
\begin{eqnarray}
\left\{
\begin{array}{rl}
F_{tr} = \sqrt{-Y/2W(Y)} \;\; , \;\;\;P_{tr}= \sqrt{-W(Y)Y/2} \vspace{0.2cm} \\  
 r = \left(\frac{\displaystyle{2Q^2}}{\displaystyle{-Y W^{3}(Y)}}\right)^{1/4} \;\;\;\;\;\;\;\;
\end{array} \right.
   \label{F+PradialYn}
\end{eqnarray}
Notice that for electric fields $Y\leq 0$. The energy-momentum components of this electrostatic solution can be obtained from the expression for $T_\mu^\nu$ in \eqref{EinsteinEqsGeneralYn} as:
  \begin{equation}
T_r^r=T_0^0 = -\frac{1}{4}Y + \frac{3n-2}{4n}\; \gamma Y^n \;\;\; , \;\;\;  T_\theta^\theta= T_\varphi^\varphi = \frac{1}{4}Y + \frac{n-2}{4n}\; \gamma Y^n ,
\label{E-M-TensorYnModelAll_n}
   \end{equation}
and their $r$-dependence can be obtained exactly as in Eq.  \eqref{F+PradialYn}. We observe that the energy density is positive definite as long as $(-1)^n \gamma >0$ which is the assumption we will take from now on. 

This solution \eqref{F+PradialYn} has the same form in flat spacetime as well. For further details we refer the reader to Ref. \cite{Verbin:2025PNLED} and note here only that the electric field does not diverge as $r\rightarrow 0$. For $n=2$, $F_{tr}(0)$ is finite, while for all $n\geq 3$ it is zero. The total field energy is always finite.

For the next stages of the discussion it will be much more convenient to transform to dimensionless quantities: the electric components will be rescaled with a parameter $\mathfrak{E}$ which has units of electric field, defined by $|\gamma| = 1/\mathfrak{E}^{2(n-1)}$ and the quantities with dimensions of length like $r$, $M$ and $Q$ will be rescaled by the length parameter  $\ell = 1/\sqrt{\kappa \mathfrak{E}^{2}}$, such that $\varrho=r/\ell$ ,  $m=M/\ell$ and $q = \kappa \mathfrak{E}\, Q$.  Within this dimensionless framework we will write the  dimensionless, static, spherically symmetric metric in terms of the single free (lapse) function $f(\rho)$ as 
\begin{equation}
ds^{2}= f(\rho) \, dt^{2}-\frac{1}{f(\rho)} \, d\rho^{2}-\rho^{2}\left(d\theta^{2}+\sin^{2}\theta \, d\phi^{2}\right) .
\label{line-element}
\end{equation}
For such a metric, Einstein's equations reduce to a single first-order equation:
\begin{equation}
\frac{1}{\rho^2 }\frac{d}{d\rho}\left[\rho(1-f)\right]=K(\rho)
\quad \Rightarrow \quad
\frac{dm}{d\rho} = \frac{\rho^2}{2} K(\rho),
\label{EinsteinEq00}
\end{equation}
where the dimensionless mass function $m(\rho)$ is defined by $f(\rho)=1-2m(\rho)/\rho$, and $K(\rho)$ is the energy density in dimensionless form. For the PINLED $Y^n$ black holes discussed here, the solution for the lapse function is written in a parametric representation with $y=-Y/\mathfrak{E}^2$ as a parameter:
\begin{eqnarray}
\left\{
\begin{array}{rl}
f(y)= 1 - \frac{2 m(y)}{\rho(y)} \;\;\;\;\;\;\;\; \vspace{0.2cm} \\ 
\rho(y)=\left( \frac{2 q^2}{y(1+y^{n-1})^3} \right)^{1/4} ,
\end{array} \right.
   \label{BHmetricYn}
\end{eqnarray}
\begin{figure}[b!]
\centering
\begin{subfigure}{0.33\textwidth}
    \includegraphics[width=\linewidth, height=\linewidth]{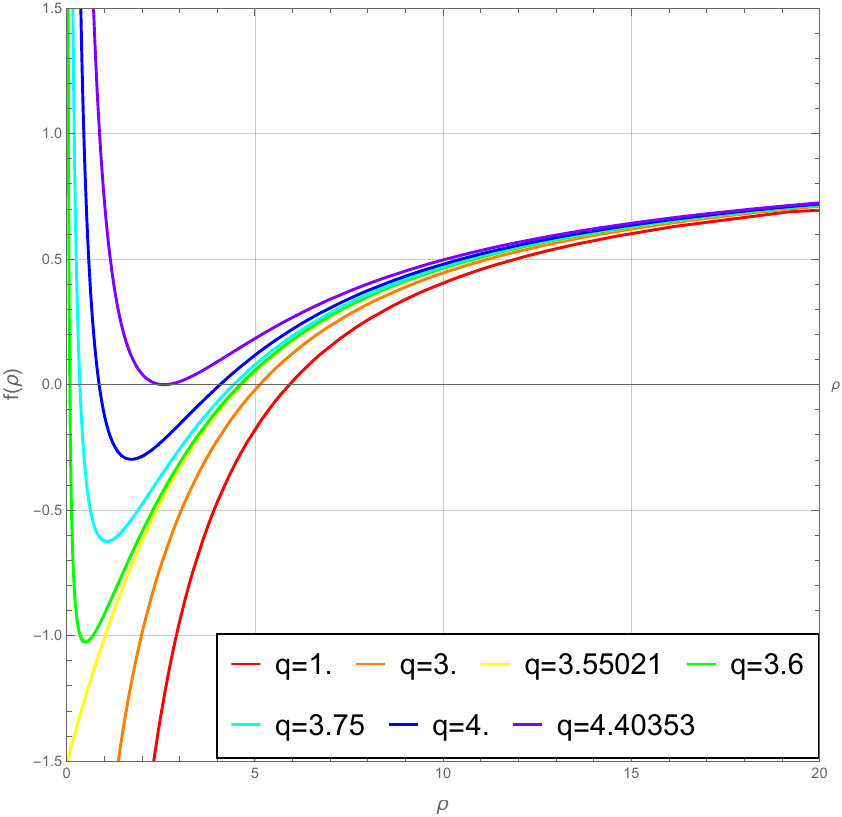} 
    \caption{$n=2$}
    \label{fig:Lapsen2SVG}
\end{subfigure}%
\begin{subfigure}{0.33\textwidth}
    \includegraphics[width=\linewidth, height=\linewidth]{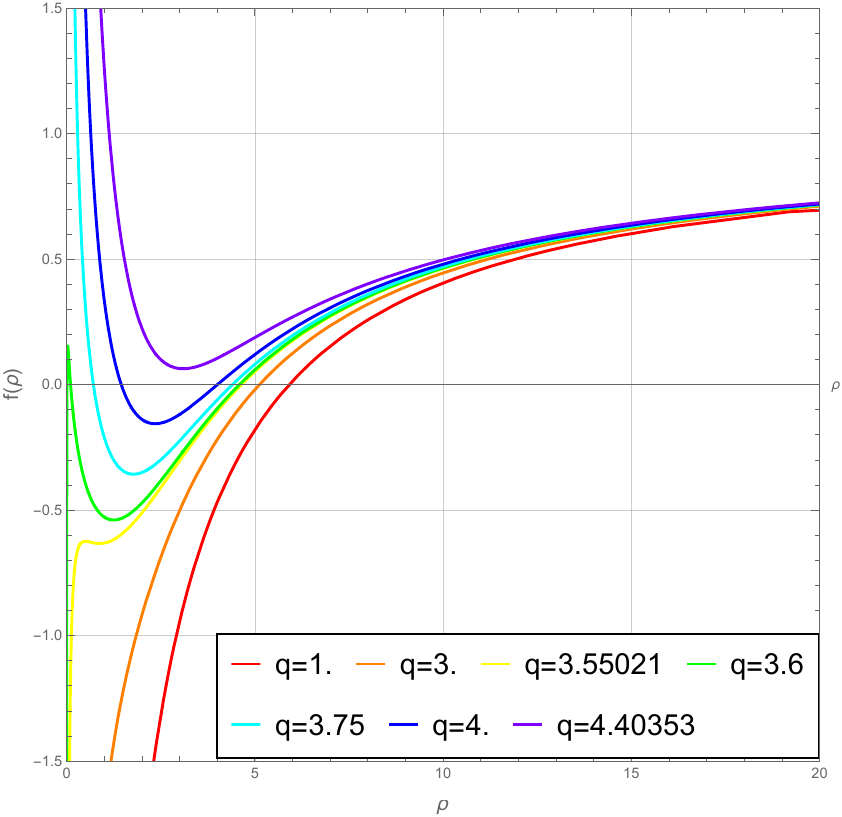}
    \caption{$n=3$}
    \label{fig:Lapsen3SVG}
\end{subfigure}%
\begin{subfigure}{0.33\textwidth}
    \includegraphics[width=\linewidth, height=\linewidth]{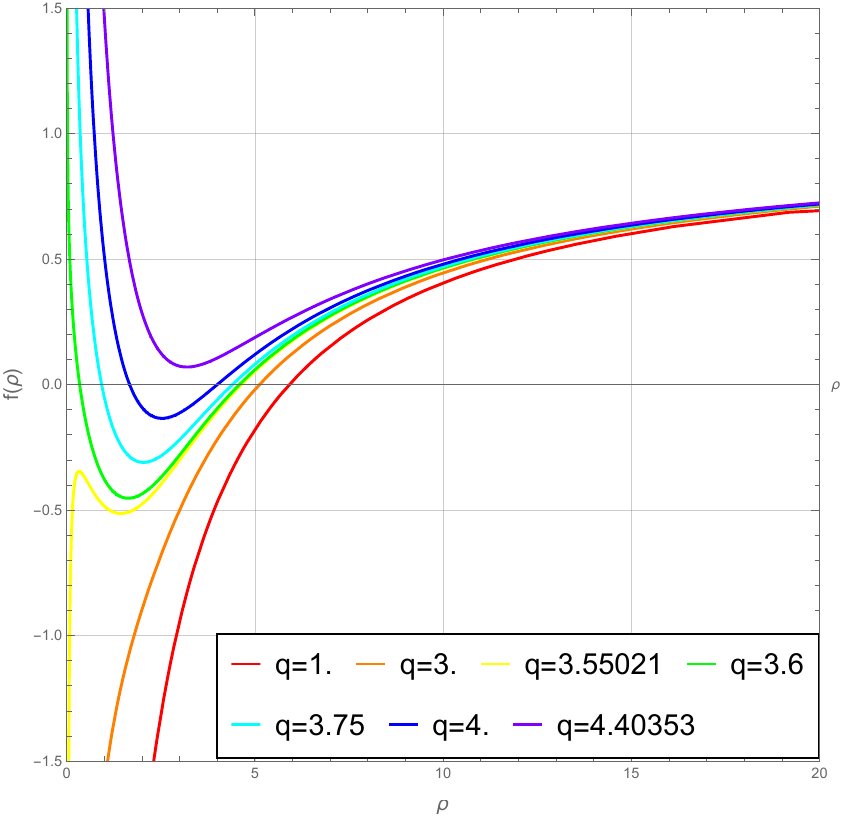} 
    \caption{$n=4$}
    \label{fig:Lapsen4SVG}
\end{subfigure}
\caption{Lapse function $f(\rho)$ of BH solutions of the PINLED $Y^n$ model  with various values of $q$ for $n$ values $n=2,3 , 4$ with $m_{BH}=3$.}
\label{fig:Lapsen}
\end{figure}
after we also transformed the second equation of \eqref{EinsteinEq00} to an equation for $m(y)$ using the explicit expression for the dimensionless energy density $ K(y) $: 
\begin{equation}
K(y)=\frac{y}{4}+\frac{3n-2}{4n} y^n
\label{DimensionlessT00}.
\end{equation} 
 The solution for the mass function is given by
\begin{eqnarray}
m(y) &=& \frac{q^{3/2}}{2^{1/4} \cdot 15(n-1)} \left[ \frac{n(17n-49) + \big(10+n(27n-101)\big)y^{n-1} - 32ny^{2(n-1)}}{4n(1+y^{n-1})^{9/4}} y^{1/4} \right. \nonumber \\
&& \left. + \frac{8}{y^{(n-2)/4}} F \left( \tfrac{1}{4}, \tfrac{n-2}{4(n-1)}, \tfrac{5n-6}{4(n-1)}, -\tfrac{1}{y^{n-1}}  \right) \right] + m_{\text{BH}}-\bar{m}_{\text{field}},
\label{BHmassFunctionYn}
\end{eqnarray}
where $F(a,b,c,z)$ is the hypergeometric function, $m_{\text{BH}}$ is the black hole mass and $\bar{m}_{\text{field}}$ is the (finite) total accumulated field mass in all space. This field mass is expressed as
\begin{eqnarray}
   \bar{m}_{\text{field}}=\frac{2^{3/4}q^{3/2}}{3}\cdot\frac{\Gamma\!\left(\tfrac{4n-3}{4(n-1)}\right)\Gamma\!\left(\tfrac{5n-6}{4(n-1)}\right)}{\Gamma\!\left(\tfrac{9}{4}\right)}.
\label{BHfieldEnergyYn}
\end{eqnarray}
In Fig.~\eqref{fig:Lapsen} and \eqref{fig:Horizonn}, one can see the profiles of the lapse function and the horizon $\rho_H$  as a function of $m_{\text{BH}}$ with different values of $q$ and $n$. PINLED $Y^n$ black holes exhibit Schwarzschild-like (S-like) solutions (with one horizon) for small charges, and Reissner–Nordström-like (RN-like) solutions (with two horizons), for larger charges.  The lower branches of the RN-like curves of Fig. \eqref{fig:Horizonn} correspond to the inner horizons. The more relevant to this paper is the event horizon which is the exterior if the number of horizons is larger than 1. Thus we will usually refer to the event horizon simply as ``horizon'' and take some care when distinction is needed. For larger $n$, the critical charge leading to naked singularities decreases. As expected, for fixed BH mass the charge $q$ always reduces the horizon radius, while for fixed charge the mass increases the horizon radius.

\begin{figure}[hbt!]
\centering
\begin{subfigure}{0.33\textwidth}
    \includegraphics[width=\linewidth, height=\linewidth]{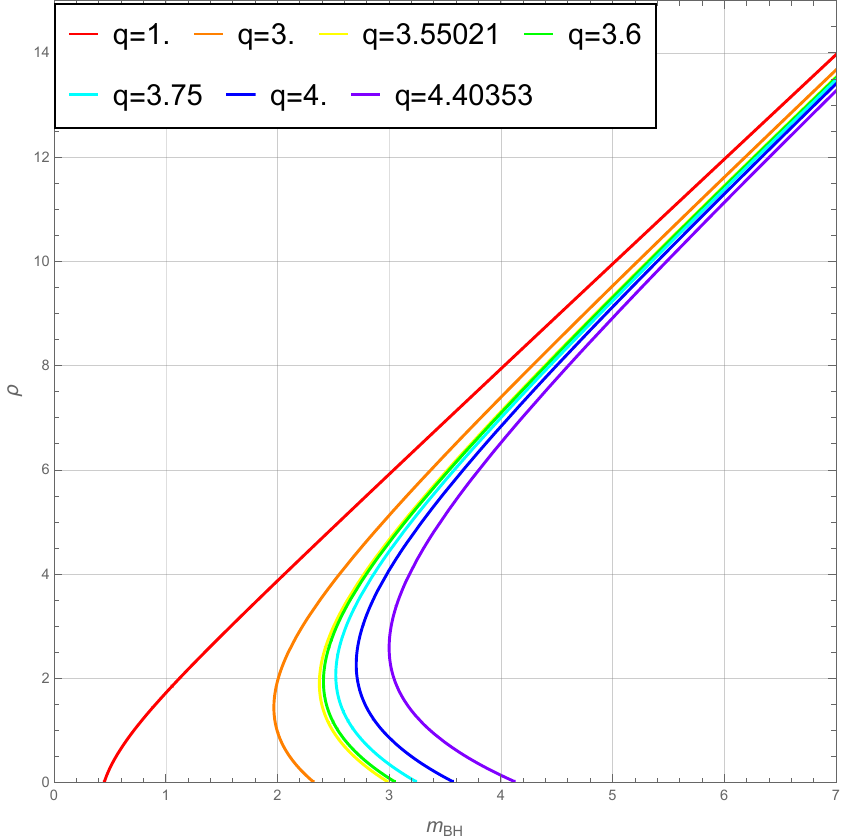} 
    \caption{$n=2$}
    \label{fig:Horizonn2SVG}
\end{subfigure}%
\begin{subfigure}{0.33\textwidth}
    \includegraphics[width=\linewidth, height=\linewidth]{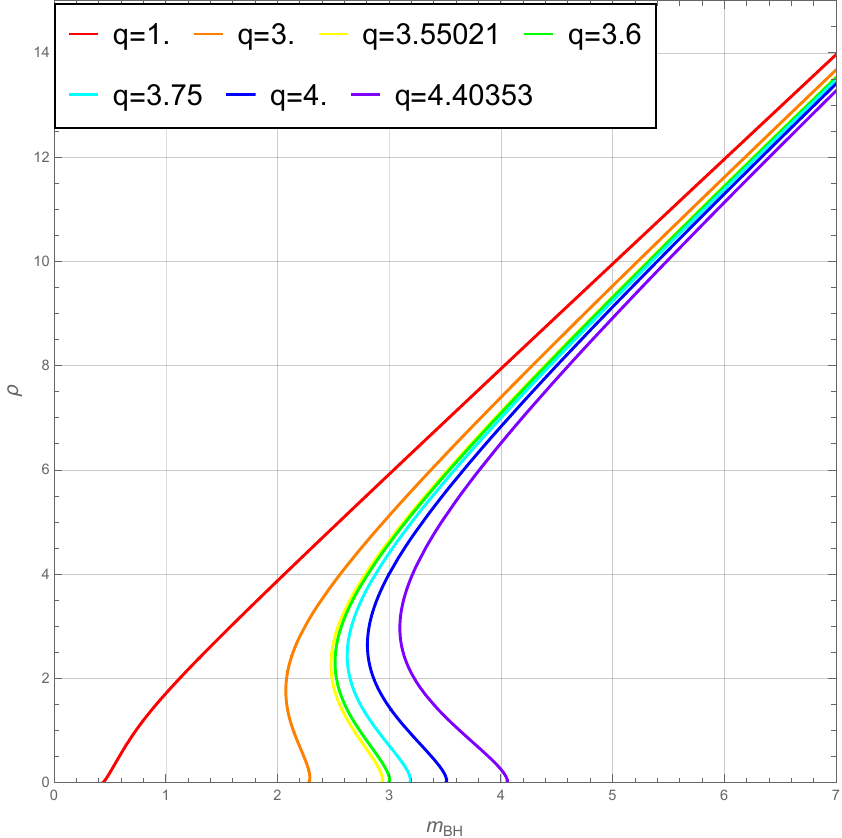}
    \caption{$n=3$}
    \label{fig:Horizonn3SVG}
\end{subfigure}%
\begin{subfigure}{0.33\textwidth}
    \includegraphics[width=\linewidth, height=\linewidth]{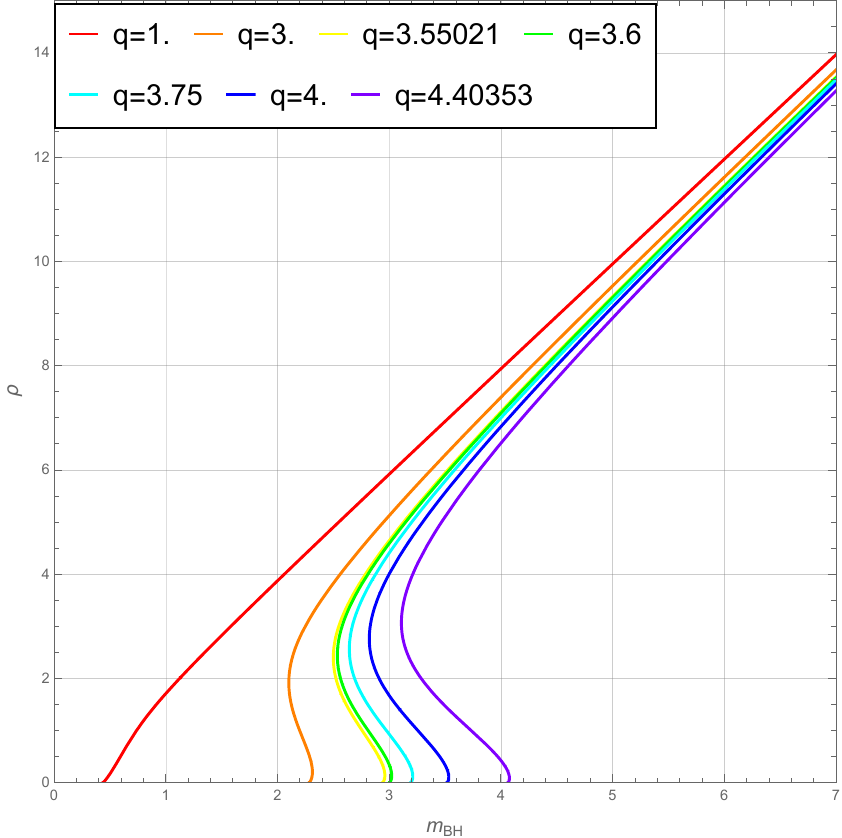} 
    \caption{$n=4$}
    \label{fig:Horizonn4SVG}
\end{subfigure}
\caption{Horizon radius $\rho_H$ versus black hole mass $m_{BH}$ for $Y^n$ BHs  with various values of $q$ for $n= 2,3 , 4$. Notice that $q=4.40353$ is the critical $q$ for $m_{BH}=3$ and $n=2$, which corresponds to a naked singularity for $n=3$ and $4$.}
\label{fig:Horizonn}
\end{figure}

\section{ Timelike and null geodesics around $Y^{n}$ black holes}
\label{timelikeandnullgeodesics}
For the study of timelike and null geodesics around the $Y^n$ BHs 
 we will follow the standard procedure by first, writing the geodesic equation and identifying the constants of motion in the equatorial plane $\theta=\pi/2$. To this end, we define the Lagrangian as $\mathcal{L}_p=\Lambda/2$, where
\begin{equation}
\Lambda =-g_{\mu\nu}\dot{x}^\mu \dot{x}^\nu
=-f(\rho) \dot{t}^2+\frac{\dot{\rho}^2}{f(\rho)}+\rho^2 \dot{\phi}^2.
\label{geodesic}
\end{equation}
where the dot represents the derivative with respect to an affine parameter. This Lagrangian immediately yields two conserved quantities: the energy $E=f(\rho)\dot{t}$ and the angular momentum $L=\rho^2 \dot{\phi}$. The Lagrangian itself is also constant along geodesics, $\Lambda=-\xi$. By an affine transformation, we can choose $\xi=\{1,0\}$ for massive and massless particles, respectively. This allows us to write the radial equation in the form of a one-dimensional mechanical energy equation for a point particle:
\begin{equation}
\dot{\rho}^2+V_{\text{eff}}=E^2,
\end{equation}
where the effective potential is given by
\begin{equation}
V_{\text{eff}}=f(\rho)\left(\frac{L^2}{\rho^2}+\xi \right). \label{VeffMassive}
\end{equation}
It is also useful to use an analogous mechanical equation for the geometry of the trajectories $r(\phi)$:
\begin{equation}
\left(\frac{d\rho}{d\phi}\right)^2+W_{\text{eff}}(\rho)=0,
\label{MechEq-Geom}
\end{equation}
where the new effective potential is
\begin{equation}
W_{\text{eff}}=\rho^2 f(\rho) + \frac{\rho^4}{L^2}\left(\xi f(\rho) -E^2\right). 
\label{Weff-Geom}
\end{equation}
Notice the structural difference of this mechanical equation with respect to the previous one: the particle energy is an additional parameter in the effective potential, while the effective energy always vanishes.

Now let us investigate massive and massless cases.
\subsection{Massive Neutral Particles ($\xi=1$)}
\label{massiveneutral}
First, we discuss the simplest types of orbits around the black hole, namely timelike circular orbits. For these, the radial velocity must vanish, $\dot{\rho}(t)=0$ for all $t$, which corresponds to $V_{\text{eff}}=E^2$ and 
$V'_{\text{eff}}(\rho)=0$. Stable circular orbits exist if $V''_{\text{eff}}(\rho)>0$ and the limiting case of this family of orbits is the \textit{innermost stable circular orbit} (ISCO), which is determined by the additional condition of marginal stability, $V''_{\text{eff}}(\rho)=0$. From $V'_{\text{eff}}(\rho)=0$ we obtain the following explicit condition for circular orbits: 
\begin{equation}
V'_{\text{eff}}(\rho )=\frac{L^2 \left(1-3 f(\rho )-\rho ^2K(\rho )\right)}{\rho ^3}
+\frac{\xi \left(1-f(\rho)-\rho ^2K(\rho )\right)}{\rho}=0,
\label{ISCOEq1}
\end{equation}
where we have used the field equation \eqref{EinsteinEq00} for eliminating $f'(\rho)$ from the equation. 

For the stability condition  we need also the second derivative 
\begin{equation}
V''_{\text{eff}}(\rho )=\frac{L^2 \left(12 f(\rho)-\rho ^3 K'(\rho )+4 \rho ^2 K(\rho )-6\right)}{\rho ^4}
-\frac{\xi \left(2 (1-f(\rho))+\rho ^3 K'(\rho )\right)}{\rho ^2}\, ,
\label{Veffpp}
\end{equation}
which as mentioned above should be positive. The ISCO is the limiting case of marginal stability which occurs for $V''_{\text{eff}}(\rho )=0$ in addition to \eqref{ISCOEq1}. 
Solving Eq.~\eqref{ISCOEq1} for the angular momentum (which is possible only for $\xi \ne 0$) and substituting in \eqref{Veffpp}, we obtain a single equation that determines the ISCO radius $\rho_i$: 
\begin{equation}
V''_{\text{eff}}(\rho_i)= 2 \xi \frac{  f(\rho_i ) \left(3 f(\rho_i )+\rho_i ^3 K'(\rho_i )+7 \rho_i ^2 K(\rho_i )-5\right)+2 \left(1-\rho_i ^2
   K(\rho_i )\right)^2}{\rho_i ^2 \left(-3 f(\rho_i )-\rho_i ^2K(\rho_i )+1\right)}  =0
  \label{ISCOEqFinal}
\end{equation} 
An equivalent equation can be obtained by recognizing that the ISCO radius corresponds to the minimum of the angular momentum function obtained by solving Eq.~\eqref{ISCOEq1} as mentioned above. The ISCO radius thus obtained, is independent of the orbit parameters (like $L$ and $E$). It depends only on the BH characteristics, $n$, $q$ and $m_{BH}$.

Since for our $Y^n$ BHs, the radial dependence of all quantities is not explicit, but is given in a parametric form using the additional radial relation  $\rho(y)$ of \eqref{BHmetricYn}, all the quantities in this equation are actually functions of $y$, or should be expressed as such, like $K'(\rho)=dK/d\rho$ using the explicit form \eqref{DimensionlessT00} for $K(y)$. After doing that, one solves Eq.~\eqref{ISCOEqFinal} with respect to $y_i$, the parametric value corresponding to the ISCO, and then obtain the associated ISCO radius as  $\rho_i = \rho(y_i)$.
\begin{figure}[hbt!]
\centering
\begin{subfigure}{0.33\textwidth}
    \includegraphics[width=\linewidth, height=\linewidth]{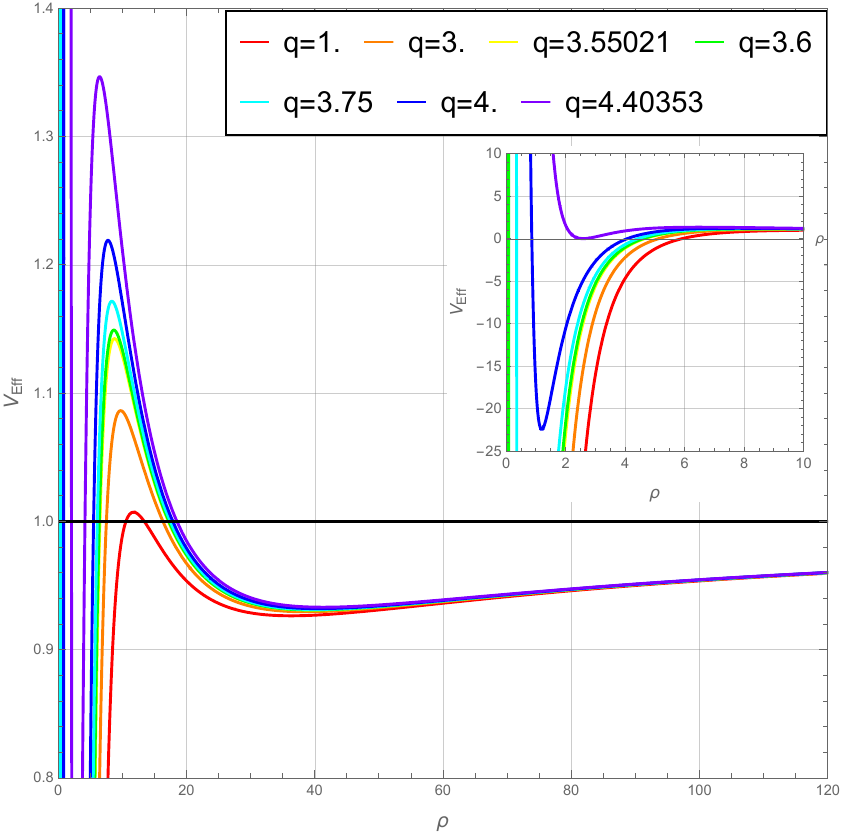} 
    \caption{$n=2$}
    \label{fig:Iscoveffn2SVG}
\end{subfigure}%
\begin{subfigure}{0.33\textwidth}
    \includegraphics[width=\linewidth, height=\linewidth]{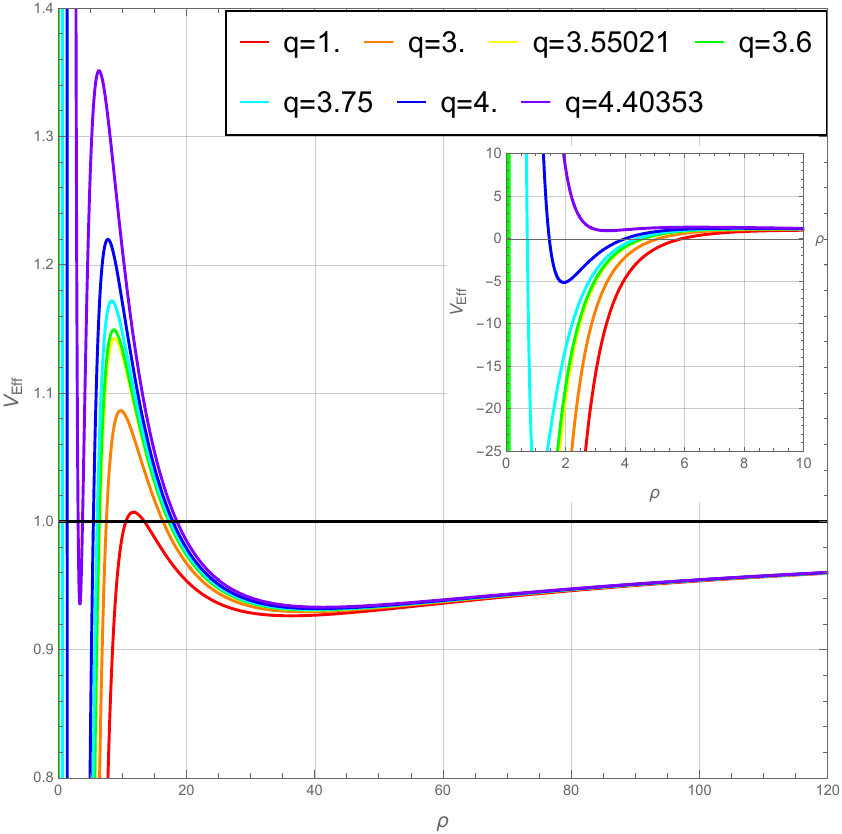}
    \caption{$n=3$}
    \label{fig:Iscoveffn3SVG}
\end{subfigure}%
\begin{subfigure}{0.33\textwidth}
    \includegraphics[width=\linewidth, height=\linewidth]{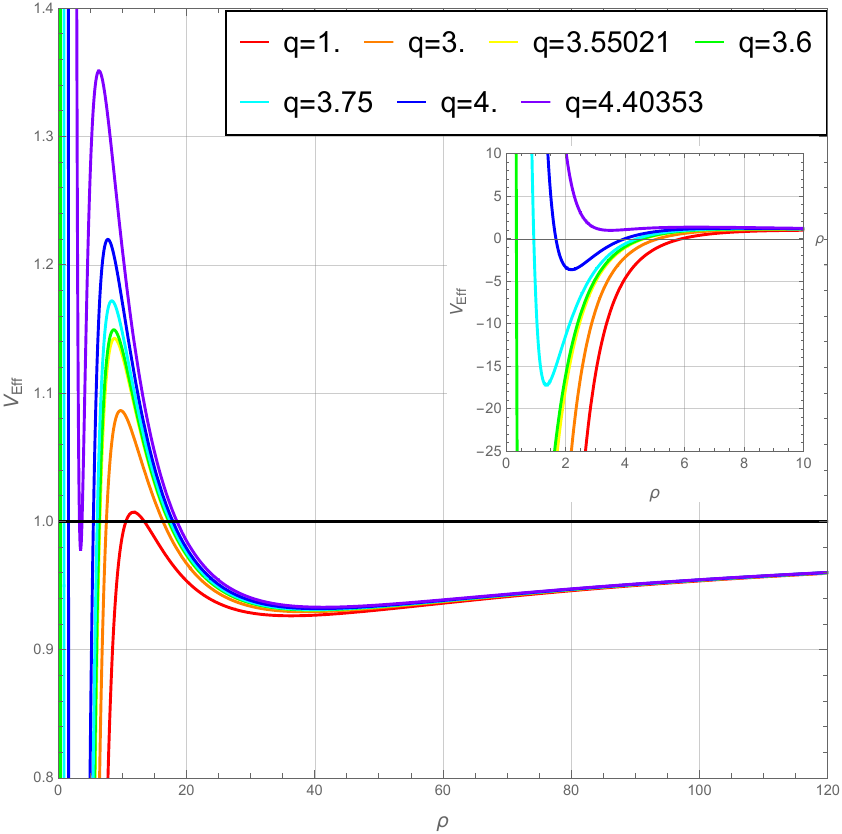} 
    \caption{$n=4$}
    \label{fig:Iscoveffn4SVG}
\end{subfigure}
\caption{Effective potential for massive particles ($\xi=1$) vs radius $\rho$ for $Y^n$ BHs  with various values of $q$ for $n$ values 2,3 and 4 with $m_{BH}=3$ and $L=4 m_{BH}$. Notice that the next to largest zero of $V_{\text{eff}}(\rho)$ is at the event horizon, and that the minimum of $V_{\text{eff}}(\rho)$ which is relevant to the circular orbits and ISCO is the shallow one further to the right around $\rho=40$ in these plots.}
\label{fig:Iscoveffn}
\end{figure}

In Fig.~\eqref{fig:Iscoveffn} we have plotted the effective potential for massive particles with a given angular momentum for different cases. We observe that depending on the nonlinearity parameter $n$ and charge $q$ one has a different nontrivial orbital structure. For low charge values (up to $q=3$) we have a Schwarzchild like behavior, a single infinitely deep potential well near the horizon and a second shallow binding potental relatively far from the horizon. Beyond these lower values, we have an intermediary state where we see the development of a double minima structure with the interior one fully inside the horizon. In this configuration there emerges a potential barrier between the inner and outer wells.  As charge increases, both potential wells become shallower, as the  black hole's grip on photons weakens since repulsive electromagnetic effects dominates.

\begin{figure}[t!]
\centering
\begin{subfigure}{0.33\textwidth}
    \includegraphics[width=\linewidth, height=\linewidth]{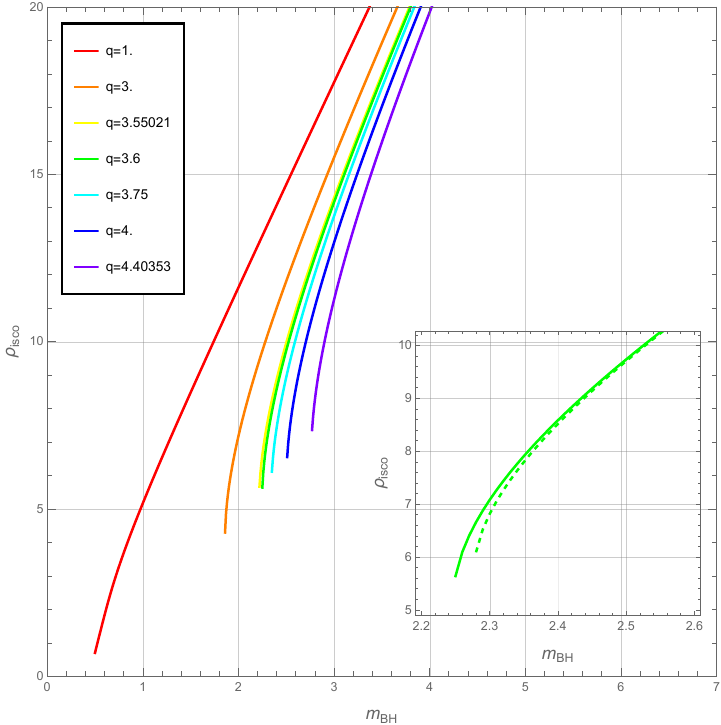} 
    \caption{$n=2$}
    \label{fig:riscovsmn2SVG}
\end{subfigure}%
\begin{subfigure}{0.33\textwidth}
    \includegraphics[width=\linewidth, height=\linewidth]{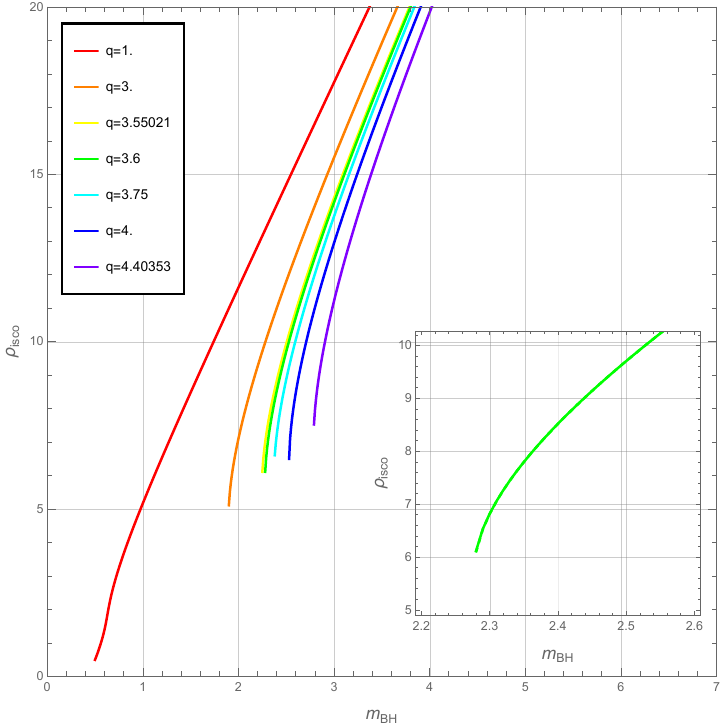}
    \caption{$n=3$}
    \label{fig:riscovsmn3SVG}
\end{subfigure}%
\begin{subfigure}{0.33\textwidth}
    \includegraphics[width=\linewidth, height=\linewidth]{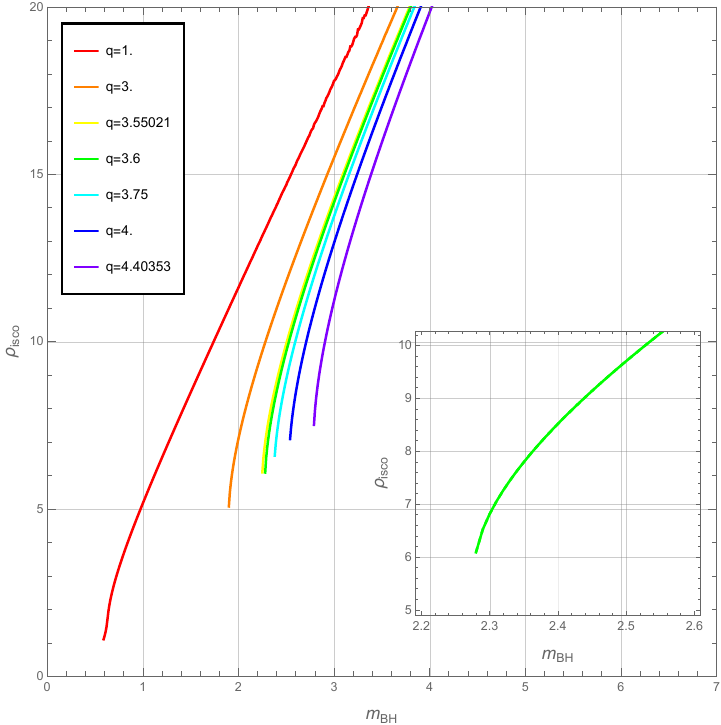} 
    \caption{$n=4$}
    \label{fig:riscovsmn4SVG}
\end{subfigure}
\caption{ISCO radius for massive particles ($\xi = 1$) as a function of the BH mass $m_{\rm BH}$ for the  $Y^n$ BHs, shown for various values of the charge $q$ and for $n = 2, 3,$ and $4$. In the inset panel, dashed curve correspond to the Reissner–Nordström (RN) case. For $n = 3$ and $4$, the difference between the PINLED and RN results becomes indistinguishable even in smaller scales.}
\label{fig:riscovsmn}
\end{figure}
Furthermore, in Figs.~\ref{fig:riscovsmn} and \ref{fig:riscovsqn} we present the behavior of the ISCO radius as a function of the  BH mass and charge, respectively. The overall trends can be read off directly from the figures, or from the numerical results summarized in Table~\ref{tab:isco} for small $q$ and $m_{\rm BH}$.  
Increasing $m_{\rm BH}$ at fixed $q$ shifts 
 $\rho_{\rm ISCO}$ outward, consistent with the expectation that a more massive black hole extends the range of its gravitational influence. Similarly, increasing $q$ at fixed $m_{\rm BH}$ shifts the ISCO inward, suggesting that in the PINLED framework the electromagnetic contribution to gravity effectively diminishes  the net attractive potential, driving  $\rho_{\rm ISCO}$ to 
smaller values.  
 We compare the PINLED results with the Reissner--Nordstr\"{o}m (RN) limit where for the comparison we write the RN lapse as
\begin{equation}
f_{\rm RN}(y)=1-\frac{2m_{\rm BH}}{\rho(y)}+\frac{q^2}{2\,\rho^2(y)} ,
\end{equation}
\begin{figure}[b!]
\centering
\begin{subfigure}{0.33\textwidth}
    \includegraphics[width=\linewidth, height=\linewidth]{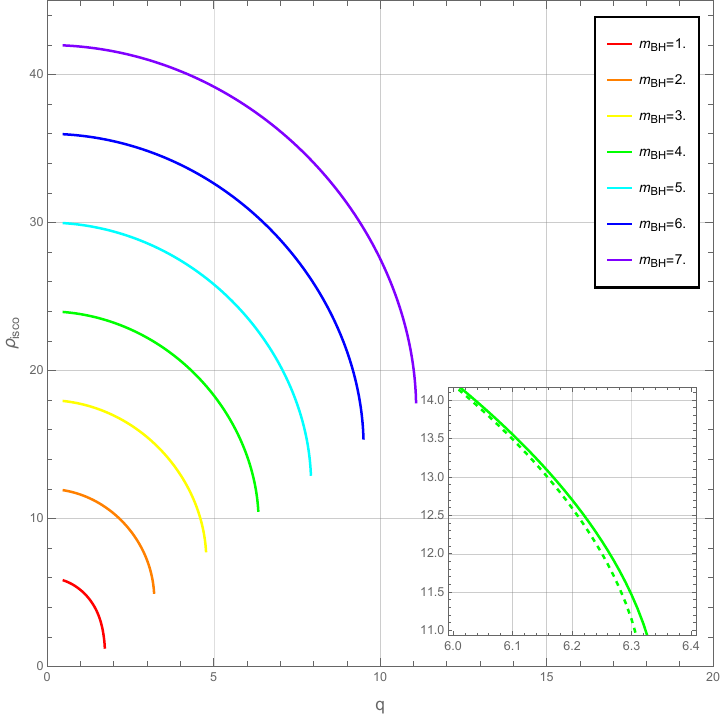} 
    \caption{$n=2$}
    \label{fig:riscovsqn2SVG}
\end{subfigure}%
\begin{subfigure}{0.33\textwidth}
    \includegraphics[width=\linewidth, height=\linewidth]{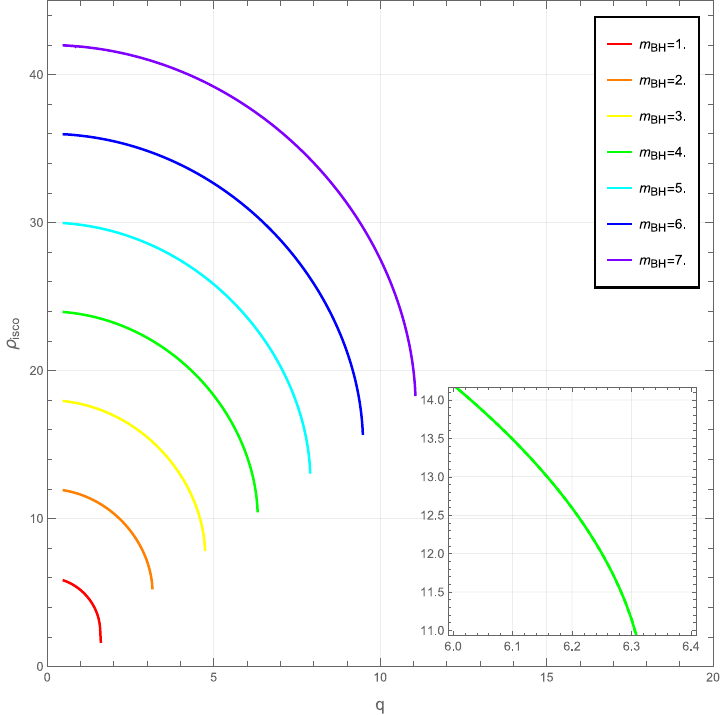}
    \caption{$n=3$}
    \label{fig:riscovsqn3SVG}
\end{subfigure}%
\begin{subfigure}{0.33\textwidth}
    \includegraphics[width=\linewidth, height=\linewidth]{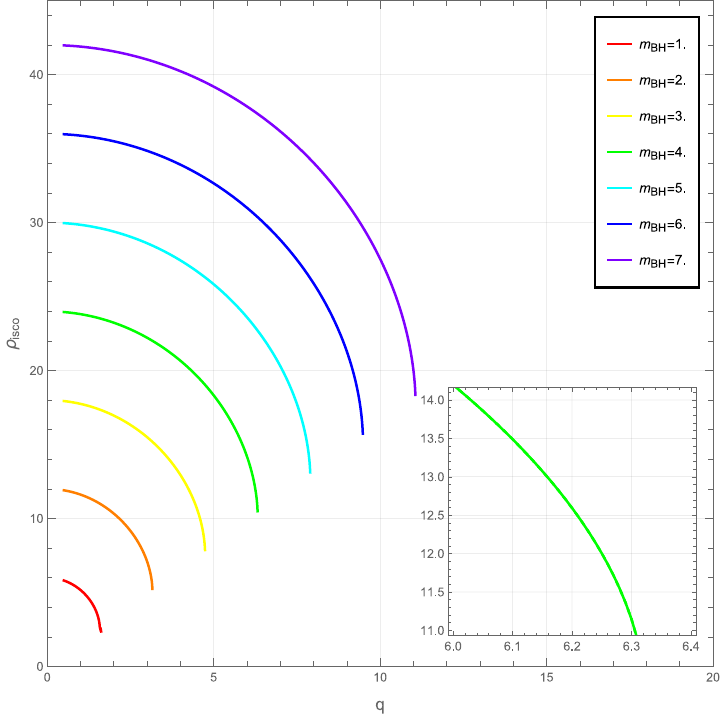} 
    \caption{$n=4$}
    \label{fig:riscovsqn4SVG}
\end{subfigure}
\caption{ISCO radius for massive particles ($\xi=1$) as a function of the charge $q$ for the $Y^n$ BHs with various fixed values of the BH mass $m_{\rm BH}$ and for $n=2,3,$ and $4$. In the inset panel the dashed curve represent the corresponding RN results.}
\label{fig:riscovsqn}
\end{figure}
using the same parametric representation of the radial coordinate $\rho(y)$ defined in Eq.~\eqref{BHmetricYn}. For the RN case we employ the same ISCO criterion based on Eq.\eqref{ISCOEqFinal} with $f_{\rm RN}(y)$ given above and $K_{\rm RN}(y)=y/4$.

We find that the two families of ISCO curves are very close to each other, with deviations being most pronounced at $n=2$ and small charge and mass, and becoming progressively smaller as $n$ increases to $3$ and $4$. By $n=3$ and $4$, the $Y^n$ curves are virtually indistinguishable from the RN ones even at finer scales, as confirmed by the inset panels in Figs.~\ref{fig:riscovsmn} and \ref{fig:riscovsqn}. This convergence 
indicates that the nonlinear corrections encoded in the $Y^n$ model become 
increasingly suppressed at higher $n$ for $\rho_{\rm ISCO}$, rendering this parameter an unfavorable probe for differentiating $Y^n$ from RN BHs. The 
results in Table~\ref{tab:isco} confirm this quantitatively: for instance, at $q=0.1$ and 
$m_{\rm BH}=0.2$, the $Y^n$ value decreases from $1.16229$ at $n=2$ to $1.16163$ at $n=4$, converging toward the RN value of $1.16163$, while at larger charges such as $q=0.3$, the deviation is more pronounced at $n=2$ but similarly converges by $n=4$.

In the next sections we will see that massless particles are more likely to tell $Y^n$ BHs apart from RN ones, since they can ``penetrate'' closer to the BH horizons, i.e. to the regions where the differences between the metric tensors of both kinds of BHs are more noticeable. Still, the deviations of the $Y^n$ BHs with respect to the RN ones will diminish with increasing charge and mass.

\begin{table}[h!]
\centering
\begin{tabular}{|c|ccc|ccc|}
\hline
 & \multicolumn{3}{c|}{RN} & \multicolumn{3}{c|}{PINLED ($n=2$)} \\
\cline{2-7}
$q$ & $m_{BH}=0.2$ & $m_{BH}=0.3$ & $m_{BH}=0.4$
    & $m_{BH}=0.2$ & $m_{BH}=0.3$ & $m_{BH}=0.4$ \\
\hline
0.1 & 1.16163 & 1.77475 & 2.38115 & 1.16229 & 1.77483 & 2.38116 \\
0.2 & 1.03329 & 1.69570 & 2.32326 & 1.04793 & 1.69729 & 2.32360 \\
0.3 & 0.709808 & 1.54994 & 2.22182 & 0.862759 & 1.56128 & 2.22392 \\
\hline
\hline
 & \multicolumn{3}{c|}{PINLED ($n=3$)} & \multicolumn{3}{c|}{PINLED ($n=4$)} \\
\cline{2-7}
$q$ & $m_{BH}=0.2$ & $m_{BH}=0.3$ & $m_{BH}=0.4$
    & $m_{BH}=0.2$ & $m_{BH}=0.3$ & $m_{BH}=0.4$ \\
\hline
0.1 & 1.16164 & 1.77475 & 2.38115 & 1.16163 & 1.77475 & 2.38115 \\
0.2 & 1.03463 & 1.69572 & 2.32326 & 1.03339 & 1.69570 & 2.32326 \\
0.3 & 0.804013 & 1.55035 & 2.22184 & 0.775924 & 1.54995 & 2.22182 \\
\hline
\end{tabular}
\caption{ISCO values for PINLED $Y^n$ black holes and RN metric for low $q$ and $m_{BH}$ range.}
\label{tab:isco}
\end{table}

\subsection{Massless Particles $\xi=0$}
\label{massless}
In this subsection, we focus on the motion of massless (neutral) particles moving along null geodesics, such as photons. This is done by setting $\xi = 0$ in Eqs.\eqref{VeffMassive} and \eqref{Weff-Geom}, which  simplifies the effective potential to
\begin{equation}
V_{\text{eff}}=f(\rho)\left(\frac{L^2}{\rho^2} \right) ,
\label{VeffLightRays}
\end{equation}
as well as the associated effective potential for light rays (i.e. $r(\phi)$) to 
\begin{equation}
W_{\text{eff}}=\rho^2 f(\rho)-\rho^4/b^2
\label{WeffLightRays}
\end{equation}
where $b=L/E $ is the impact parameter of a light ray. In what follows we will use both kinds of effective potential according to our convenience.

 Depending on the profile of the effective potential $V_{\text{eff}}(\rho)$, it can exhibit both minima and maxima corresponding to stable and unstable circular orbits, respectively. The potential for massless particles typically admits both, but outside the horizon there are  only unstable circular orbits.  These orbits occur at the photon sphere (or  photon ring in 2+1 dimensions), where the centrifugal repulsion is exactly balanced by the gravitational attraction.
\begin{figure}[hbt!]
\centering
\begin{subfigure}{0.33\textwidth}
    \includegraphics[width=\linewidth, height=\linewidth]{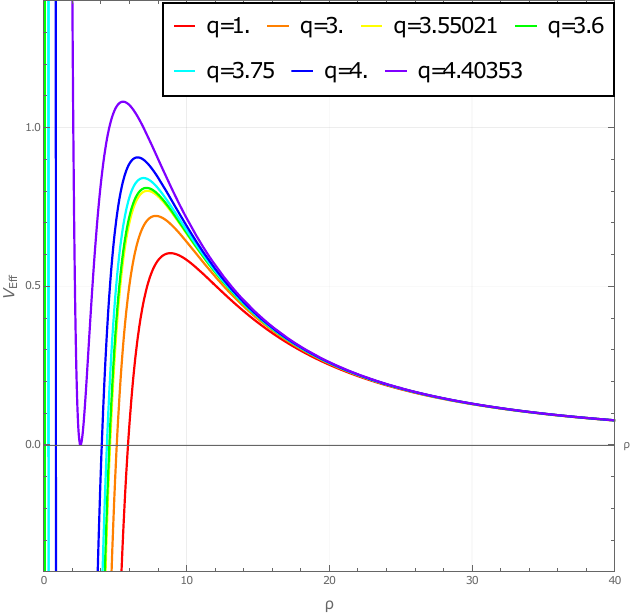} 
    \caption{$n=2$}
    \label{fig:psveffn2SVG}
\end{subfigure}%
\begin{subfigure}{0.33\textwidth}
    \includegraphics[width=\linewidth, height=\linewidth]{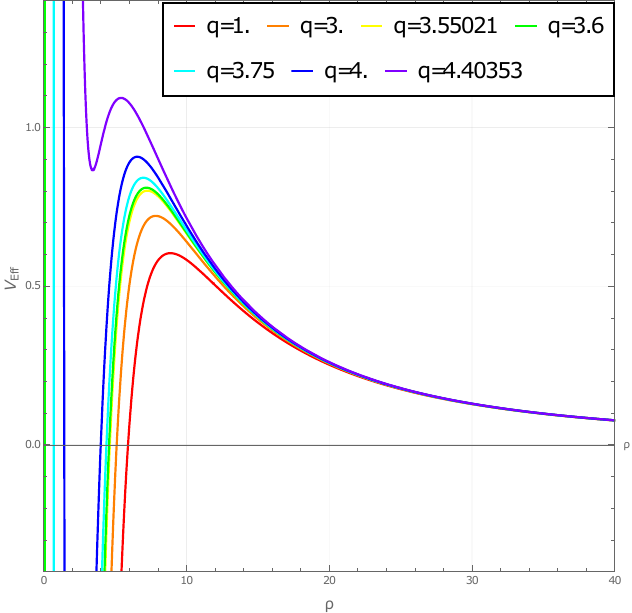}
    \caption{$n=3$}
    \label{fig:psveffn3SVG}
\end{subfigure}%
\begin{subfigure}{0.33\textwidth}
    \includegraphics[width=\linewidth, height=\linewidth]{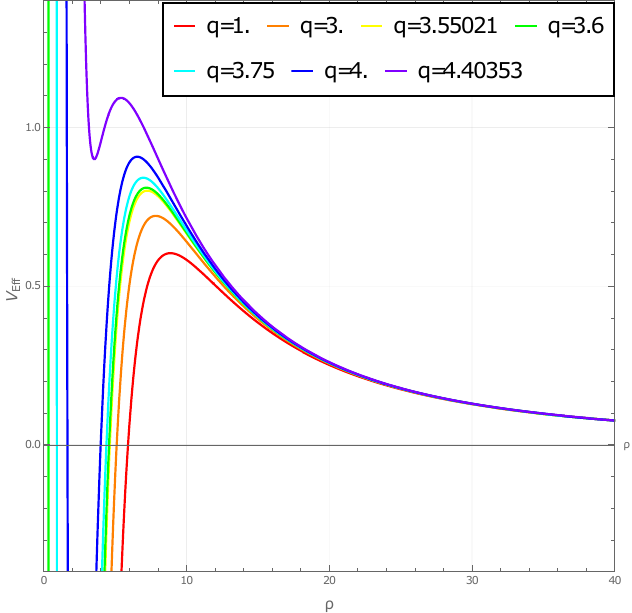} 
    \caption{$n=4$}
    \label{fig:psveffn4SVG}
\end{subfigure}
\caption{Effective potential $V_{\text{eff}}(\rho)$ for massless particles ($\xi=0$) for $Y^n$ BHs  with various values of $q$ for $n$ values 2,3 and 4 with $m_{BH}=3$. The photon spheres correspond to the static unstable $\rho(\lambda)$  ``sitting'' at the top of the potential hill.}
\label{fig:psveffn}
\end{figure}

In Fig.~\ref{fig:psveffn} we can see the profile of effective potential $V_{\text{eff}}(\rho)$ of  $Y^n$ black holes for various values of $n$ with a given angular momentum. The minima that correspond to stable photon orbits are hidden behind horizon which is located at the exterior root of $V_{\text{eff}}(\rho)$. Furthermore as expected, we observe an evolution from a Schwarzschild like profile (low $q$) to a more oscillatory profile with 
additional turning point as $q$ increases. 

The condition for circular null geodesics is obtained by requiring that the radial acceleration vanishes, which is equivalent to extremizing the effective potential $dV_{\text{eff}}/d \rho=0$ as we saw already for massive particles. For  the potential \eqref{VeffLightRays} for null geodesics this yields  
\begin{equation}
 V'_{\text{eff}}=\frac{L^2}{\rho^3}(\rho f'(\rho)-2 f(\rho)) = 0 .
\quad \Rightarrow \quad \rho_{ps} f'(\rho_{ps})=2 f(\rho_{ps})
\label{EqPhotonSphereRho1}
\end{equation}
Furthermore, using the field equation   \eqref{EinsteinEq00} we may eliminate  $f'(\rho)$ and obtain a simple algebraic relation for the photon sphere radius $\rho_{ps}$, which can be also be obtained directly from the RHS of \eqref{ISCOEq1} with $\xi = 0$:
\begin{equation}
1-3f(\rho_{ps})-\rho^2_{ps} K(\rho_{ps})=0 .
\label{EqPhotonSphereRho2}
\end{equation}
More useful for our $Y^n$ solutions which utilize parametric representation, is \eqref{EqPhotonSphereRho2} rewritten in terms of $y$, or its alternative version written in terms of $y$ and the mass function:
\begin{figure}[b!]
\centering
\begin{subfigure}{0.33\textwidth}
    \includegraphics[width=\linewidth, height=\linewidth]{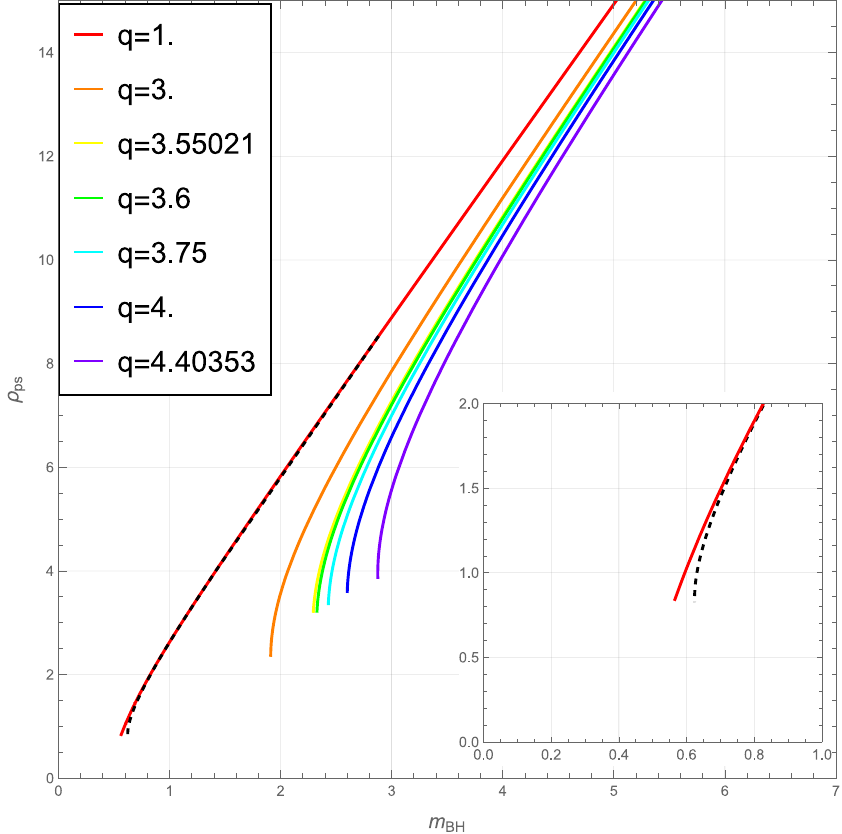} 
    \caption{$n=2$}
    \label{fig:rpsvsmn2SVG}
\end{subfigure}%
\begin{subfigure}{0.33\textwidth}
    \includegraphics[width=\linewidth, height=\linewidth]{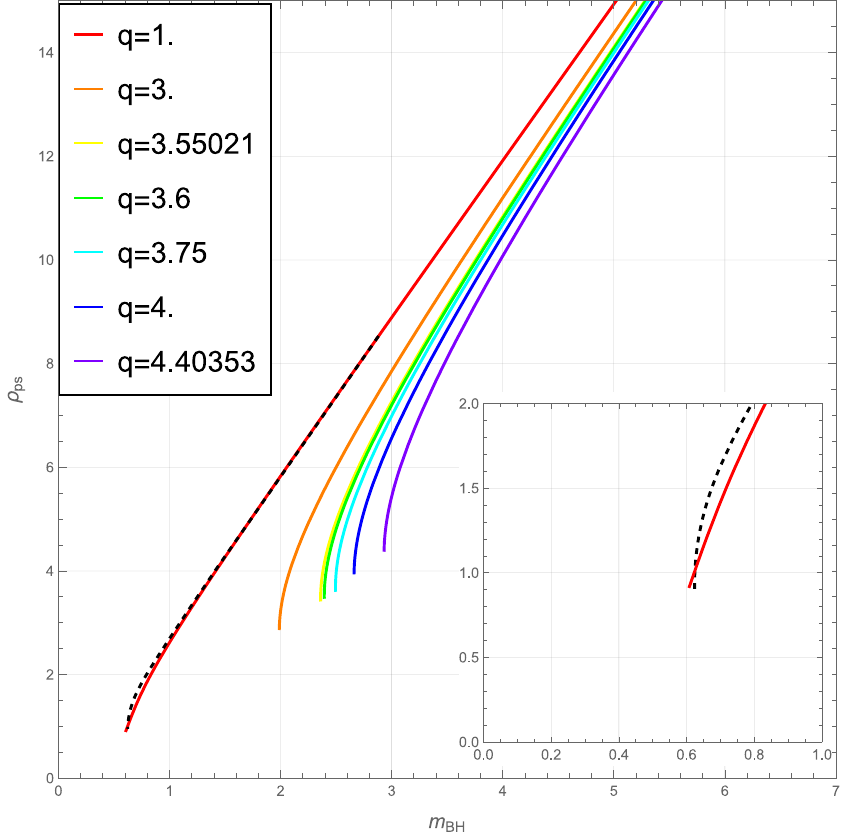}
    \caption{$n=3$}
    \label{fig:rpsvsmn3SVG}
\end{subfigure}%
\begin{subfigure}{0.33\textwidth}
    \includegraphics[width=\linewidth, height=\linewidth]{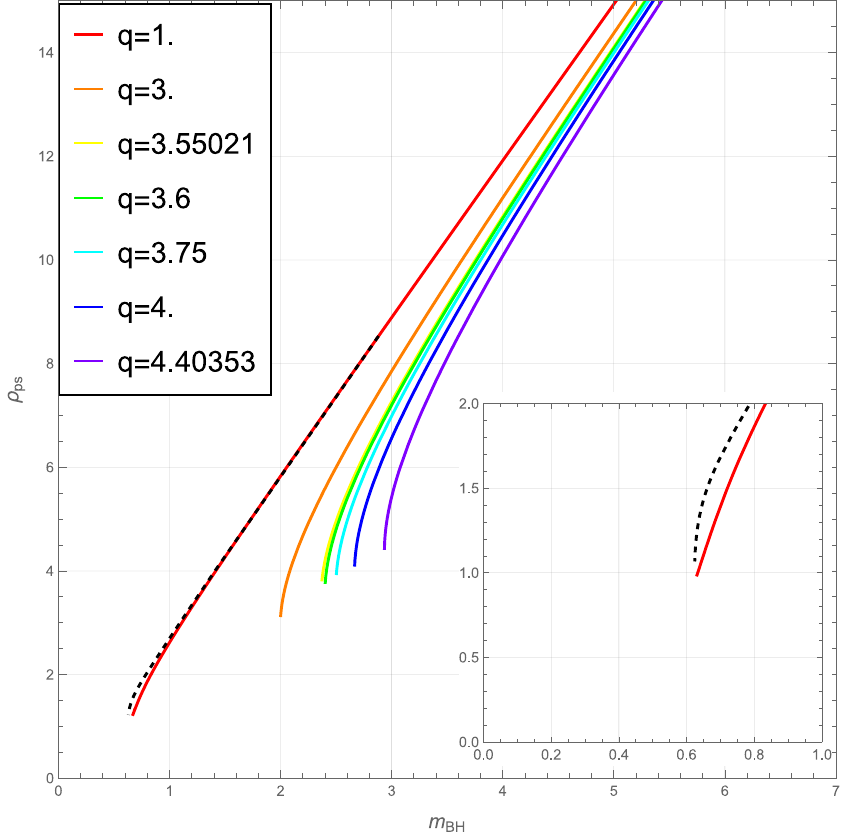} 
    \caption{$n=4$}
    \label{fig:rpsvsmn4SVG}
\end{subfigure}
\caption{Photon sphere radius $\rho_{\rm ps}$ as a function of the black hole mass $m_{\rm BH}$ for the $Y^n$ black hole with various values of the charge $q$ and for $n=2,3,$ and $4$. Black dashed curves represent the corresponding RN case with $q=1$ for each value of $n$. Other parameter choices yield similar qualitative behavior.}
\label{fig:rpsvsmn}
\end{figure}
\begin{equation}
 \frac{6 m(y_{ps})}{\rho(y_{ps})}-\rho^2 (y_{ps}) K(y_{ps})-2  =0 .
 \label{phcon}
\end{equation}
For given solutions  $y_{ps}$ of Eq.~\eqref{phcon} one can obtain the corresponding photon sphere radius $\rho_{ps}$ with the condition $\rho_{ps}>\rho_{H}$. We note that like the ISCO radius, the photon sphere radius is independent of the specific orbital parameters, but depends only on the BH characteristics, $n$, $q$ and $m_{BH}$.

In Fig.~\eqref{fig:rpsvsmn}, the dependence of the photon sphere radius $\rho_{ps}$ on the black hole mass $m_{BH}$ is displayed for various charge parameters $q$ and nonlinearity parameter $n$. The solid curves represent the $Y^n$ BH photon sphere radii, while the dashed curves which split from the lower parts of the solid curves correspond to the RN BHs for the charge values $q=1$. All  $\rho_{ps}$ profiles with fixed charge increase monotonically with $m_{BH}$. For a given $m_{BH}$, increasing the charge $q$ leads to a systematic decrease of $\rho_{ps}$ reflecting the repulsive PINLED contribution to the effective gravitational pull. The overall behavior remains qualitatively similar for different $n$, though larger $n$ slightly modifies the profiles. 

\begin{figure}[b!]
\centering
\begin{subfigure}{0.33\textwidth}
    \includegraphics[width=\linewidth, height=\linewidth]{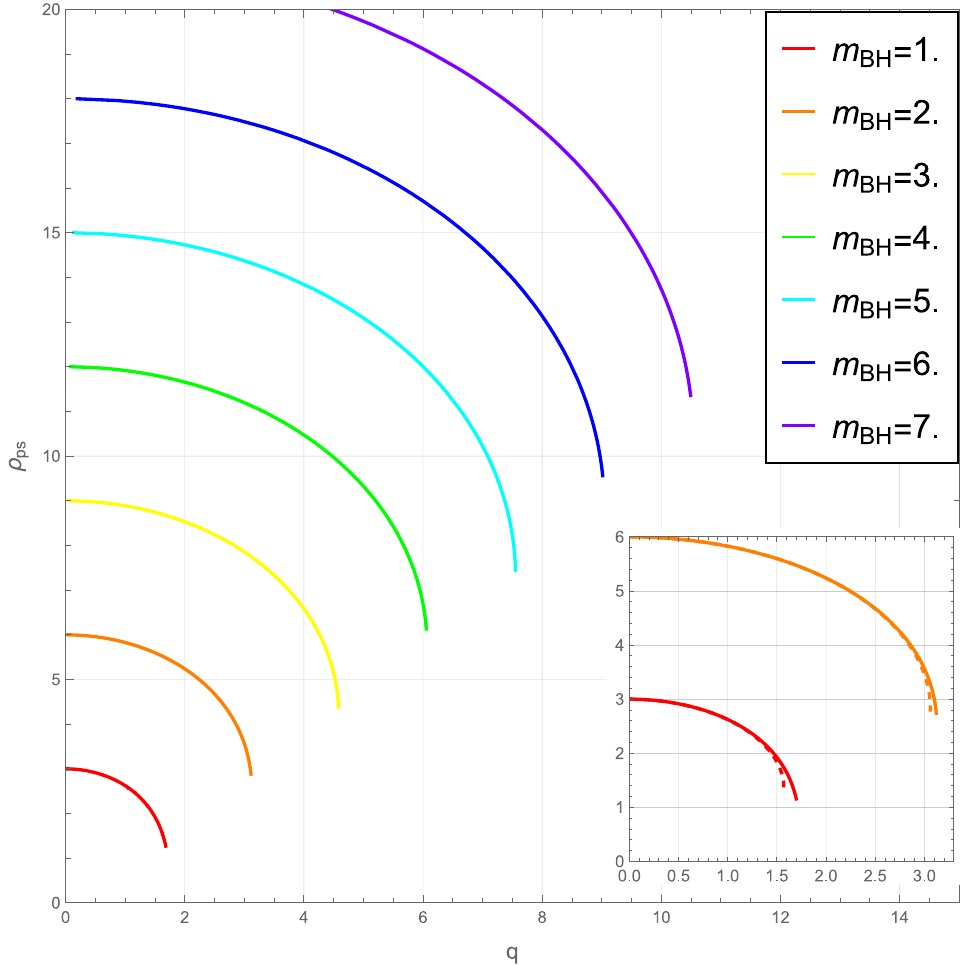} 
    \caption{$n=2$}
    \label{fig:psvsqn2}
\end{subfigure}%
\begin{subfigure}{0.33\textwidth}
    \includegraphics[width=\linewidth, height=\linewidth]{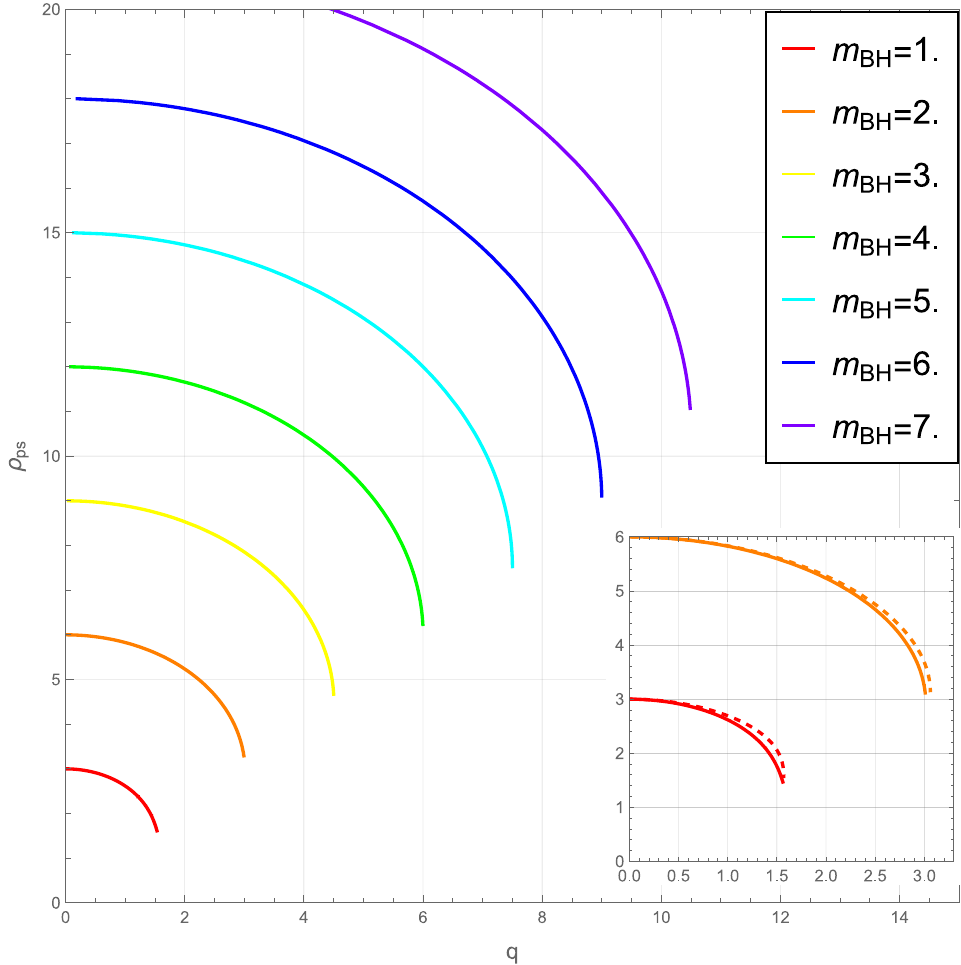}
    \caption{$n=3$}
    \label{fig:psvsqn3}
\end{subfigure}%
\begin{subfigure}{0.33\textwidth}
    \includegraphics[width=\linewidth, height=\linewidth]{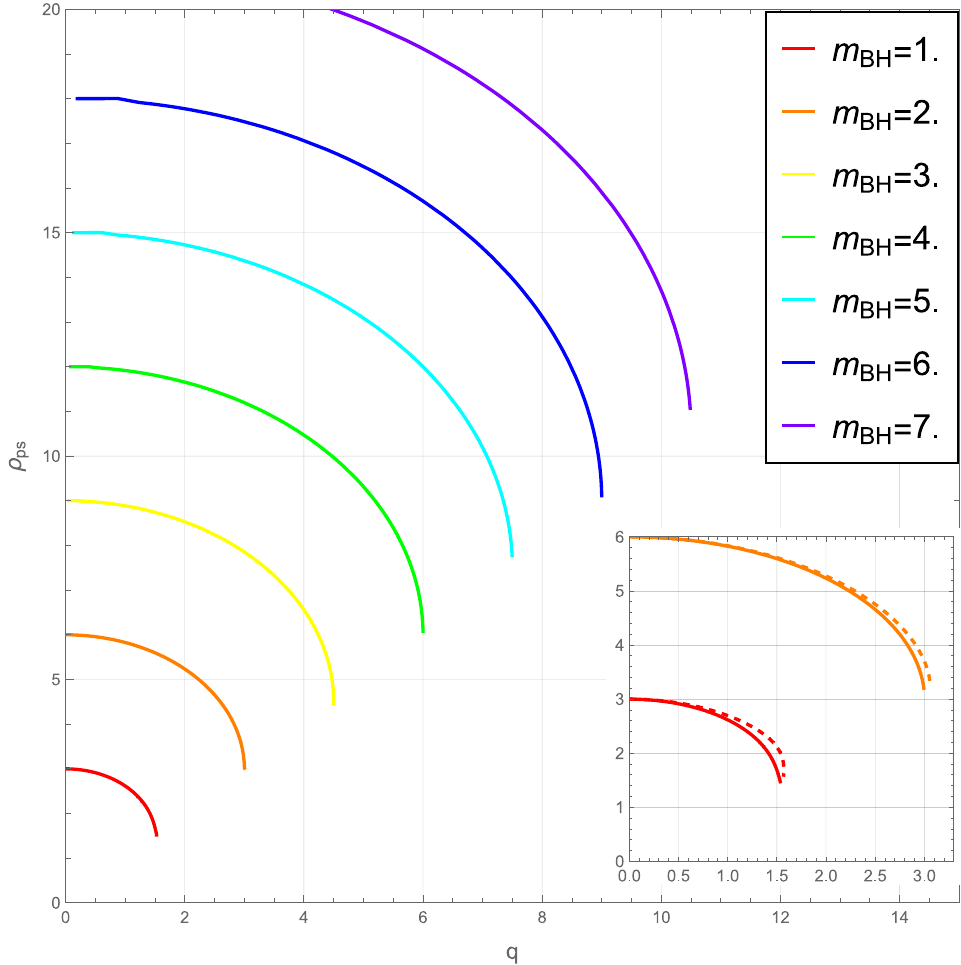} 
    \caption{$n=4$}
    \label{fig:psvsqn4}
\end{subfigure}
\caption{Photon sphere radius $\rho_{\rm ps}$ as a function of the charge $q$ for the  $Y^n$ black hole with various fixed values of the black hole mass $m_{\rm BH}$ and for $n=2,3,$ and $4$. Dashed curves represent the corresponding RN results. Other parameter choices yield similar qualitative behavior.}

\label{fig:psvsqn}
\end{figure}

 Complementarily, in Fig.~\eqref{fig:psvsqn} the variation of  $\rho_{ps}$ with $q$ is shown for several fixed black hole masses. For a given mass, $\rho_{ps}$ decreases monotonically as the charge increases. For larger masses, the photon sphere radius is correspondingly greater, consistent with the stronger gravitational influence of more massive black holes. The qualitative behavior remains similar across different $n$ values, with increasing $n$ producing only minor quantitative shifts in the curves.

In both Figs.~\eqref{fig:rpsvsmn} and \eqref{fig:psvsqn}, we observe that the photon sphere radius $\rho_{\rm ps}$ in the $Y^n$ and RN spacetimes exhibits qualitatively similar behavior over most of the parameter space. Noticeable deviations arise primarily in the low-mass and high-charge (near-extremal) regime, where the values of $\rho_{\rm ps}$ differ between the two models. In particular, for $n=2$ the RN photon sphere radius is smaller than its $Y^n$ counterpart, whereas for higher values of $n$ the Reissner--Nordström radius becomes larger in this regime.


\subsection{Shadow Radius}
\label{shadow}
In the era of very-long-baseline interferometry (VLBI) , which gave us the first images of the supermassive black holes M87$^*$ and SgrA$^*$, the concept of the black hole "shadow" has become a central observable in testing strong-field gravity. The shadow is not the event horizon itself, but a dark silhouette cast against the bright, lensed background emission of the surrounding accretion flow. This silhouette appears because photons with sufficiently small impact parameters fall into the black hole, creating a region of photon capture. The boundary of this dark region is defined by the critical photon orbits.
\begin{figure}[b!]
\begin{center}
\includegraphics[width=0.8\textwidth]{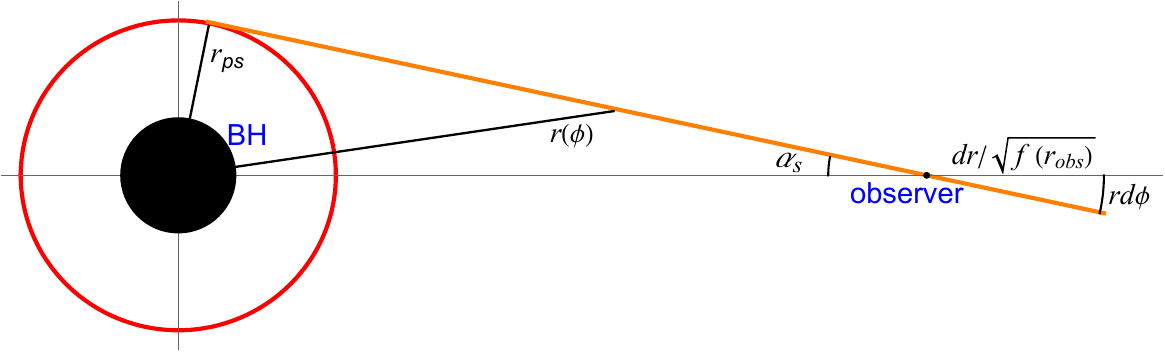}
\caption{\small{Illustration of shadow calculation from the observer’s position $r_{obs}$. The shadow angle $\alpha_{s}$ is the angle between the limiting light ray (which arrives from the photon sphere (radius $r_{ps}$) at the observer) and the radial direction. The infinitesimal angular and radial lengths are calculated at $r=r_{obs}$}.}
\label{shadow_diagram}
\end{center}
\end{figure}
\begin{figure}[b!]
\centering
\begin{subfigure}{0.33\textwidth}
    \includegraphics[width=\linewidth, height=\linewidth]{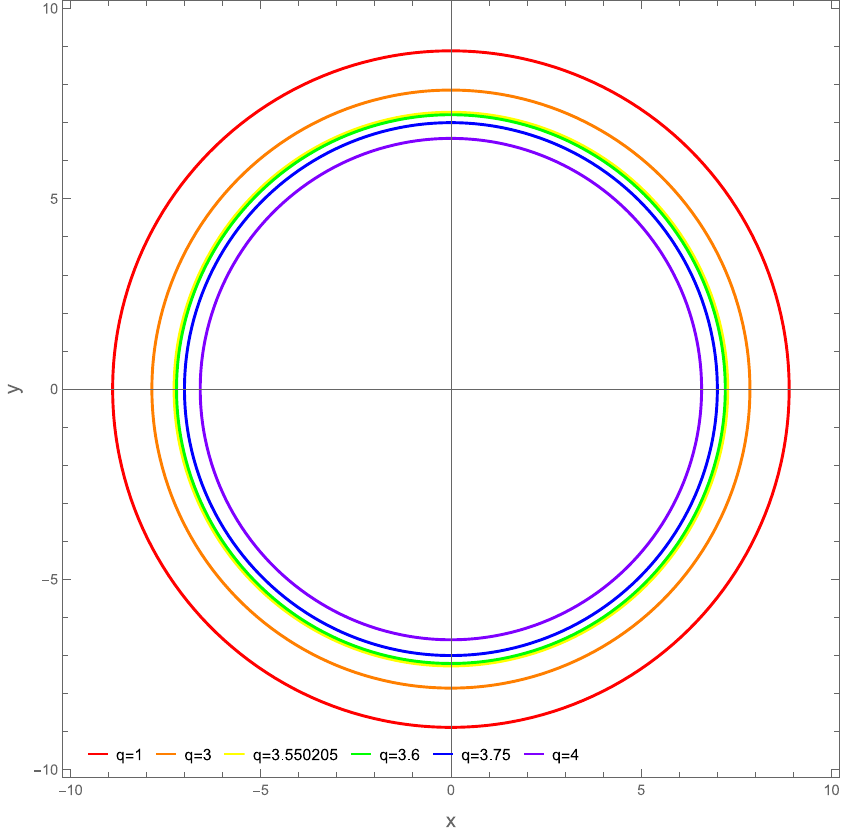} 
    \caption{$n=2$}
    \label{fig:cppsn2SVG}
\end{subfigure}%
\begin{subfigure}{0.33\textwidth}
    \includegraphics[width=\linewidth, height=\linewidth]{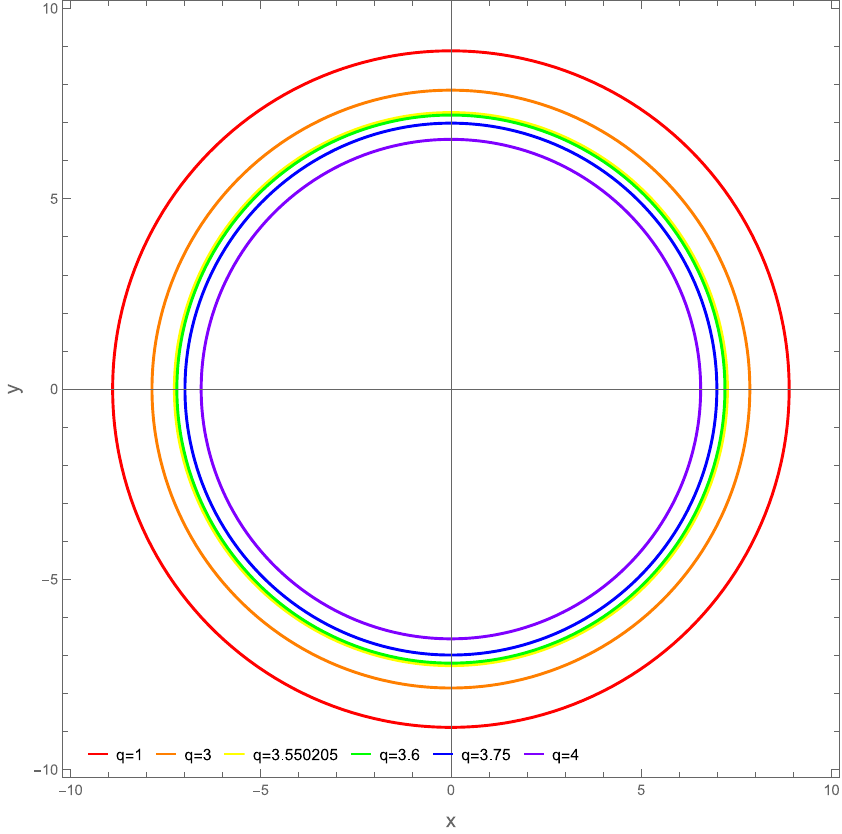}
    \caption{$n=3$}
    \label{fig:cppsn3SVG}
\end{subfigure}%
\begin{subfigure}{0.33\textwidth}
    \includegraphics[width=\linewidth, height=\linewidth]{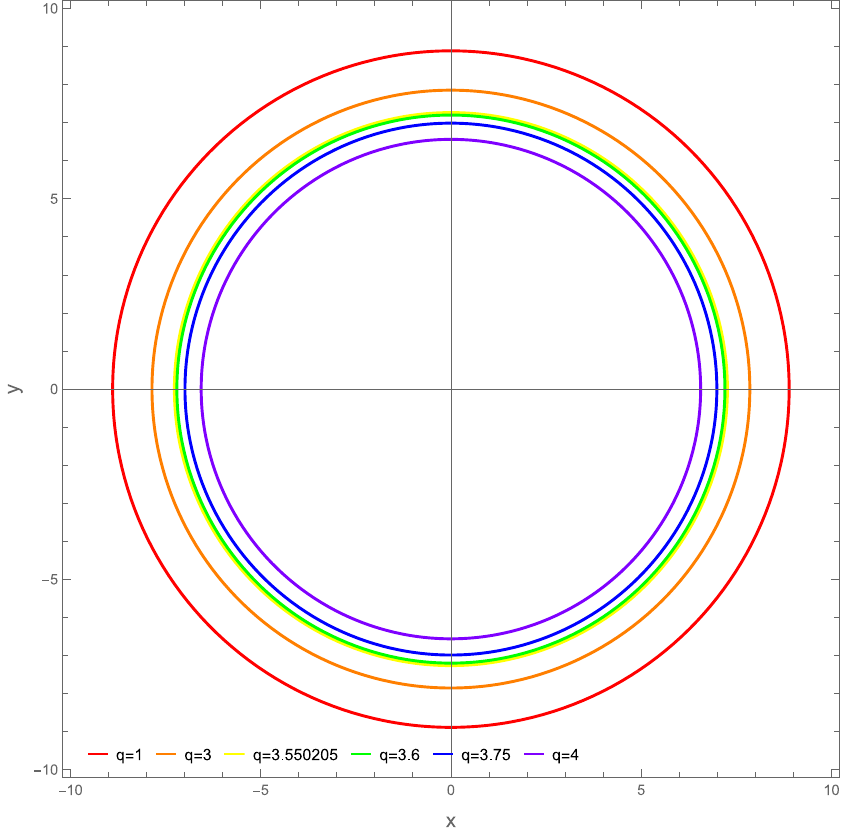} 
    \caption{$n=4$}
    \label{fig:cppsn4SVG}
\end{subfigure}
\caption{Shadow values for $Y^n$ BH  with various values of charge $q$ for $n$ values 2,3 and 4 for $m_{BH}=3$.}
\label{fig:cppsn}
\end{figure}

For a distant observer, the shadow is a circular (or nearly circular) disk whose apparent radius is governed by the innermost photon orbits. A key quantity for describing these orbits is the impact parameter  $b=L/E$ defined already at the previous subsection. As we have shown there, unstable photon circular orbits exist at a critical radius $\rho_{ps}$. The corresponding critical impact parameter $b_{crit}$ separates photons that fall into the black hole from those that escape to infinity. In asymptotically flat spacetimes, this critical impact parameter directly corresponds to the shadow radius $r_s$ observed at infinity. For our spherically-symmetric line element \eqref{line-element}, we can easily relate the shadow radius to the photon sphere quantities. For that we notice that the critical light rays  that arrive at the observer are those which connect the photon sphere and the observer represented by the ray shown in Fig. \ref{shadow_diagram}. 
The angle $\alpha$ between such a light ray and the radial direction at the location of the observer satisfies the following relation written in terms of the dimensionless radial variable: 

\begin{figure}[hbt!]
\centering
\begin{subfigure}{0.33\textwidth}
    \includegraphics[width=\linewidth, height=\linewidth]{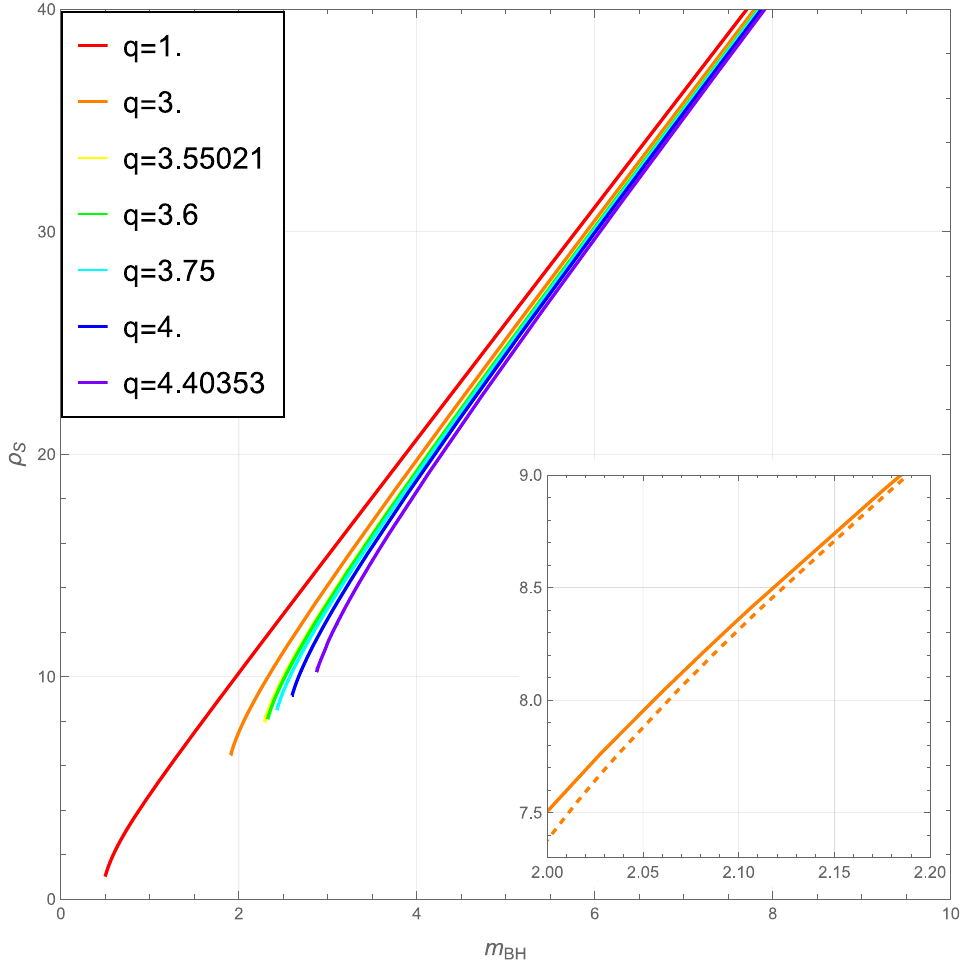} 
    \caption{$n=2$}
    \label{fig:shadowvsmn2SVG}
\end{subfigure}%
\begin{subfigure}{0.33\textwidth}
    \includegraphics[width=\linewidth, height=\linewidth]{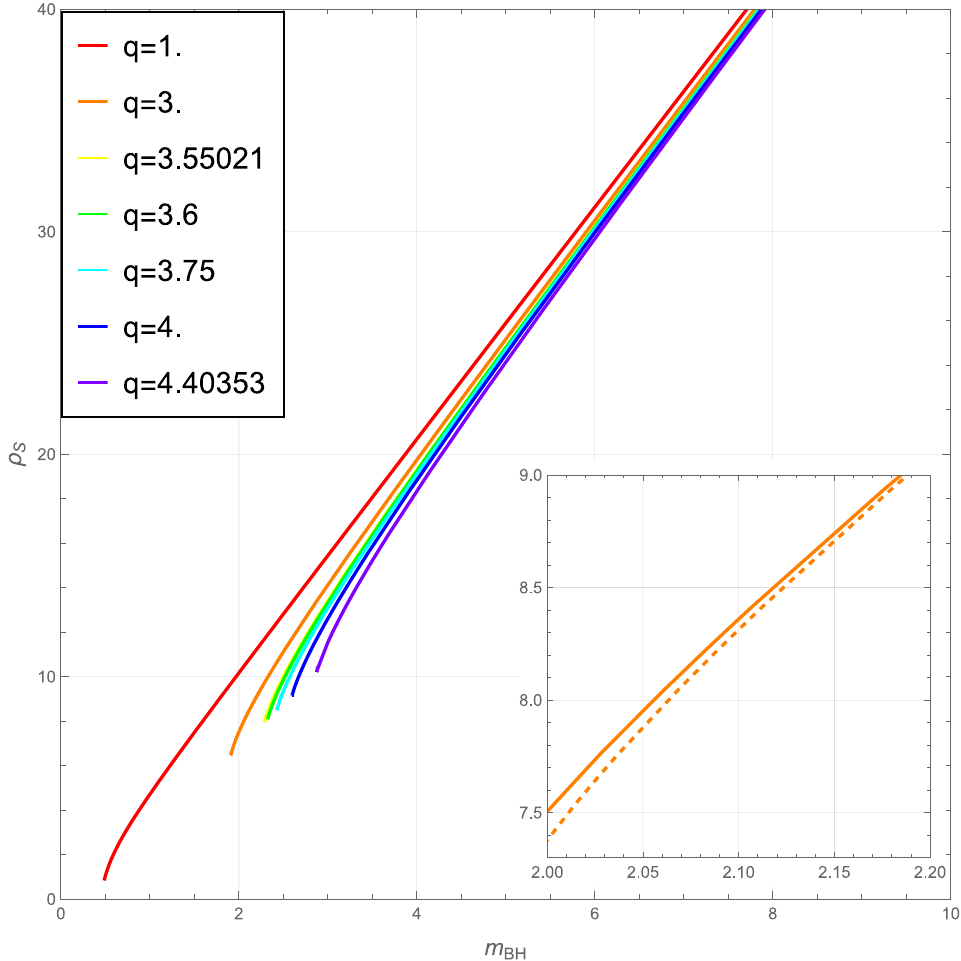}
    \caption{$n=3$}
    \label{fig:shadowvsmn3SVG}
\end{subfigure}%
\begin{subfigure}{0.33\textwidth}
    \includegraphics[width=\linewidth, height=\linewidth]{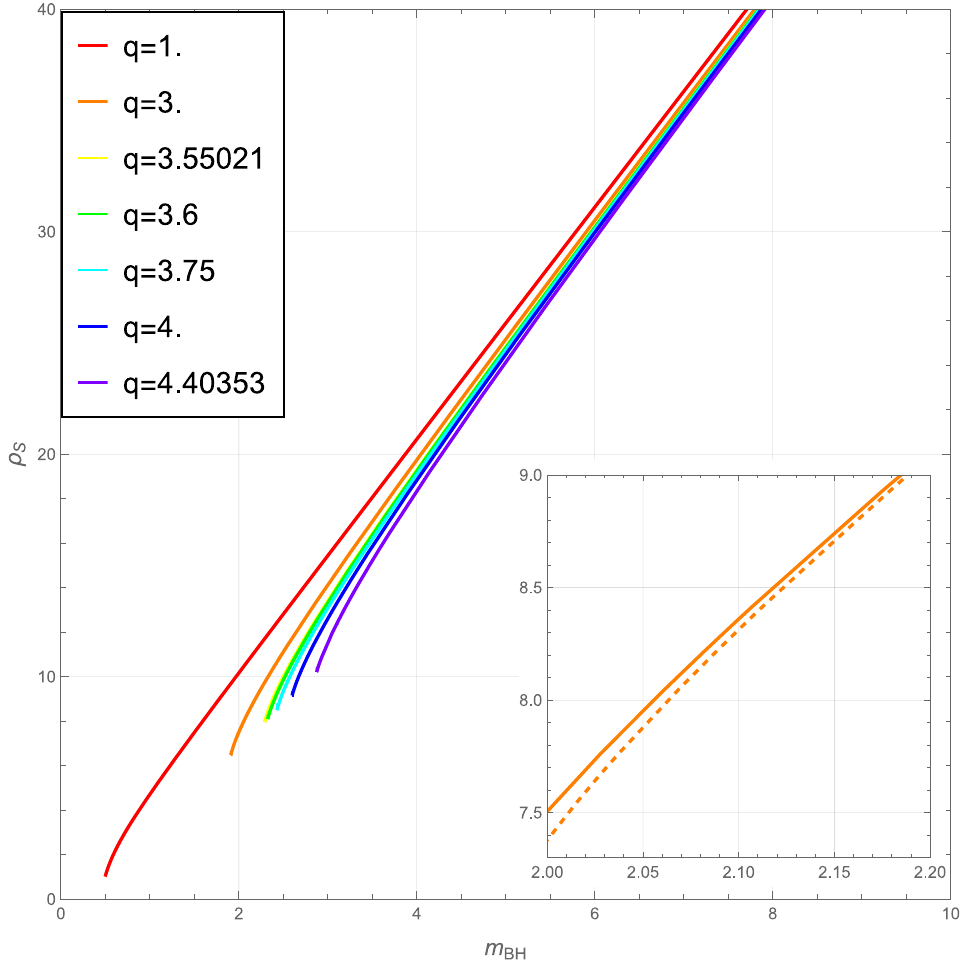} 
    \caption{$n=4$}
    \label{fig:shadowvsmn4SVG}
\end{subfigure}
\caption{Shadow radius $\rho_{s}$ as a function of the black hole mass $m_{\rm BH}$ for the $Y^n$ black hole with various values of the charge $q$ and for $n=2,3,$ and $4$. Dashed curve denote the corresponding RN result for the same charge value. Other choices of $q$ exhibit similar qualitative behavior.}
\label{fig:shadowvsmn}
\end{figure}
\begin{equation}
\cot^2{\alpha} = \left(\frac{d\rho/\sqrt{f(\rho_{obs})}}{\rho_{obs}\, d\phi}\right)^2  = \frac{1}{\rho_{obs}^2 f(\rho_{obs})} \left(\frac{d\rho}{d\phi}\right)^2 \bigg|_{\rho=\rho_{obs}} = \frac{-W_{\text{eff}}(\rho_{obs})}{\rho_{obs}^2 f(\rho_{obs})}
\label{CotAlpha1}
\end{equation}
where for the last term we used Eq.\eqref{MechEq-Geom}. Actually, this expression is still valid for any light ray which arrives at the observer. In order to specify in light rays which connect the observer with the photon sphere, we need to pick up the impact parameter $b_{crit}$ which yields $W_{\text{eff}}(\rho_{ps})=0$ and use that value in $W_{\text{eff}}(\rho_{obs})$. Looking at the relevant expression \eqref{WeffLightRays}, we find immediately that $b_{crit}^2 = \rho_{obs}^2/f(\rho_{obs})$, and by substituting in \eqref{CotAlpha1} we obtain: 
\begin{eqnarray}\label{shadow-angle-2}
\sin^2 \alpha_{s}=\frac{\rho_{ps}^2}{f(\rho_{ps})}\frac{f(\rho_{obs})}{\rho^2_{obs}}.
\end{eqnarray}
\begin{figure}[b!]
\centering
\begin{subfigure}{0.33\textwidth}
    \includegraphics[width=\linewidth, height=\linewidth]{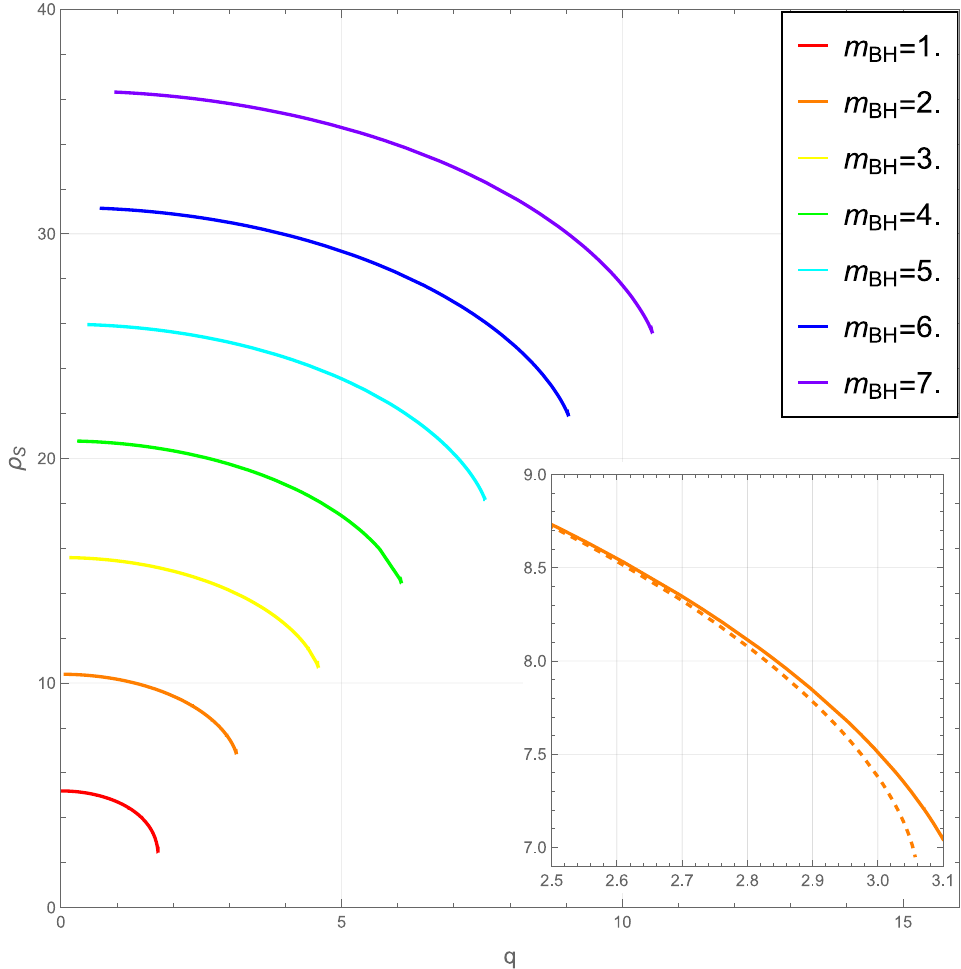} 
    \caption{$n=2$}
    \label{fig:shadowvsqn2SVG}
\end{subfigure}%
\begin{subfigure}{0.33\textwidth}
    \includegraphics[width=\linewidth, height=\linewidth]{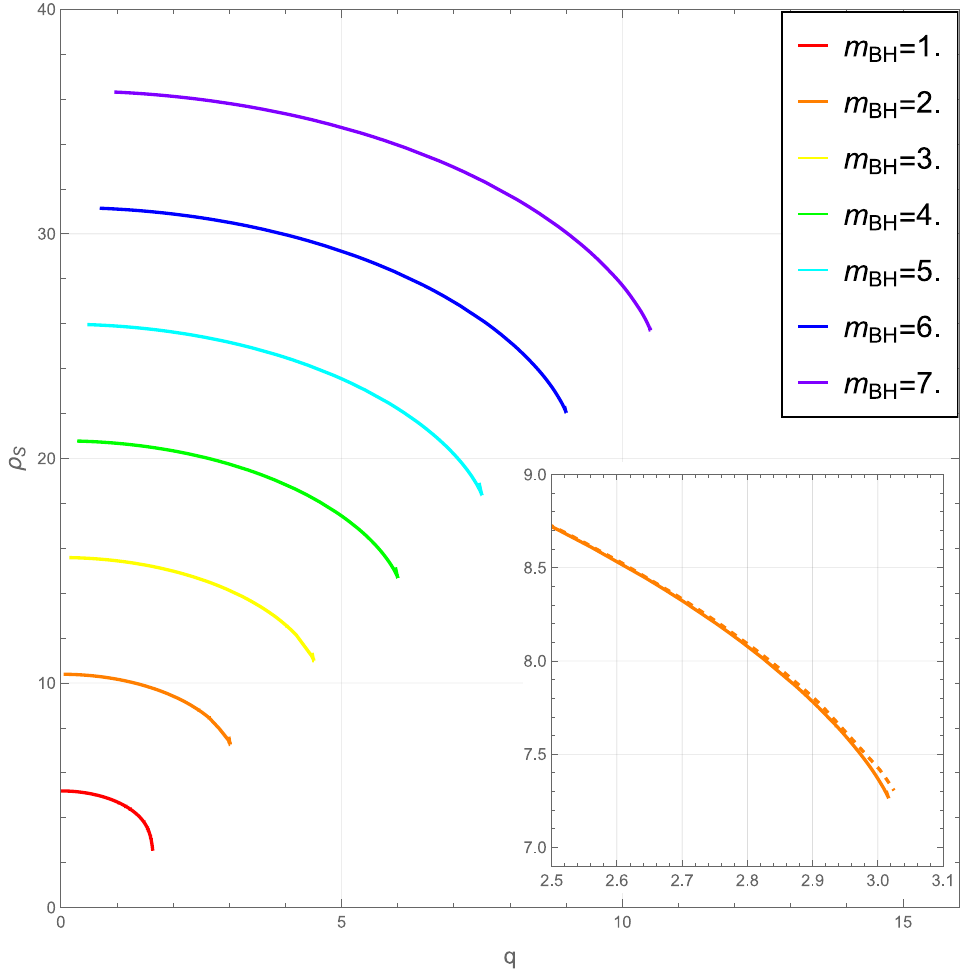}
    \caption{$n=3$}
    \label{fig:shadowvsqn3SVG}
\end{subfigure}%
\begin{subfigure}{0.33\textwidth}
    \includegraphics[width=\linewidth, height=\linewidth]{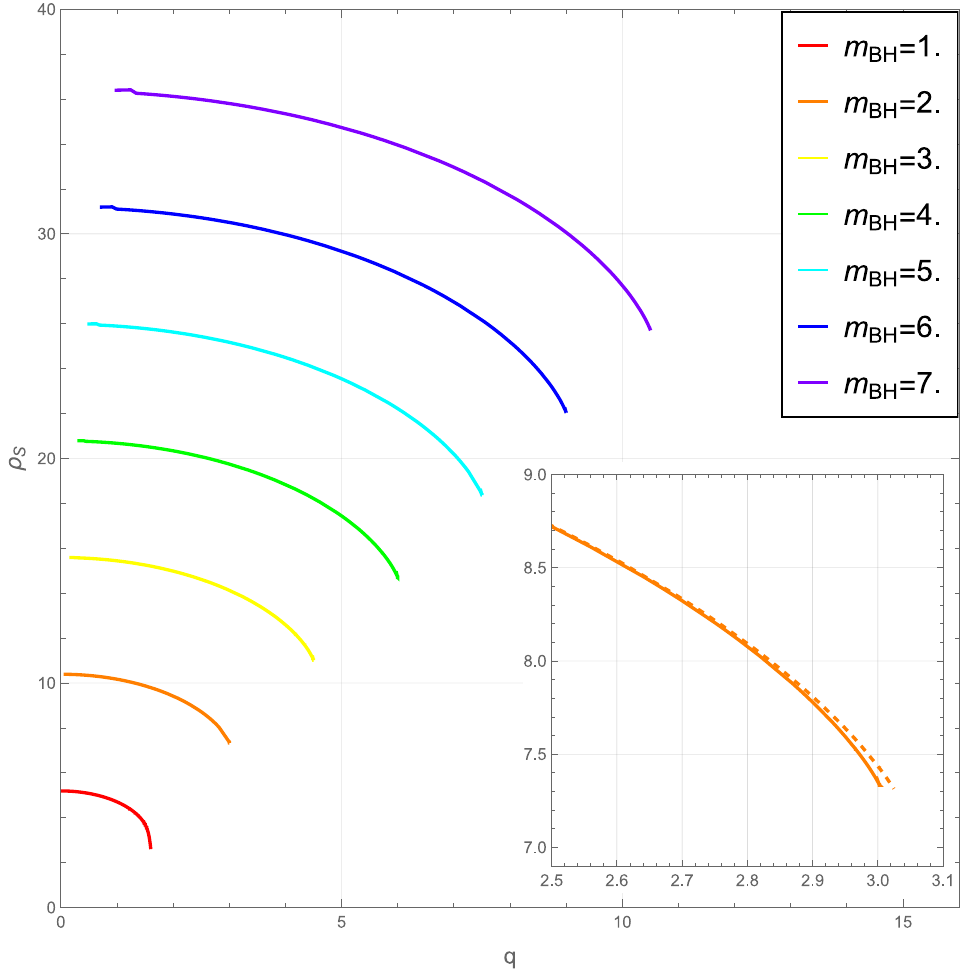} 
    \caption{$n=4$}
    \label{fig:shadowvsqn4SVG}
\end{subfigure}
\caption{Shadow radius $\rho_{s}$ as a function of the charge $q$ for various values of the black hole mass $m_{\rm BH}$ and for $n=2,3,$ and $4$. Dashed curve denote the corresponding RN result for the same mass value. Other parameter choices exhibit similar qualitative behavior.}
\label{fig:shadowvsqn}
\end{figure}
The radius of the BH shadow for an observer at a large distance can be obtained using Eq.(\ref{shadow-angle-2}) as
\begin{eqnarray}
\label{shadow-radius}
\rho_{s}&\simeq&\rho_{obs} \sin \alpha_{s} \simeq \frac{\rho_{ps}}{\sqrt{f(\rho_{ps})}}{\sqrt{f(\rho_{obs})}},
\end{eqnarray}
so the shadow radius as perceived by a static observer at infinity is given by:
\begin{equation}
\rho_s=\frac{\rho_{ps}}{\sqrt{f(\rho_{ps})}}
\end{equation}
In the context of PINLED model, using \eqref{EqPhotonSphereRho2} we obtain the expression for the shadow radius as
\begin{equation}
\rho_s=\frac{\sqrt{3}  \rho(y_{ps})}{\sqrt{1-\rho^2 (y_{ps}) K(y_{ps})}}
\end{equation}
for use in parametric representation.

Figure~\eqref{fig:cppsn} shows the calculated shadow radius for the $Y^n$ black holes as a geometric diagram. It is evident that the shadow radius decreases systematically with increasing charge $q$. Figure~\eqref{fig:shadowvsmn} illustrates the dependence of the shadow radius $\rho_s$ on the black hole mass $m_{\rm BH}$, where $\rho_s$ increases monotonically as $m_{\rm BH}$ grows. Finally, Fig.~\eqref{fig:shadowvsqn} presents the behavior of $\rho_s$ as a function of $q$ for a fixed mass, again exhibiting a decrease with increasing charge up to a critical value. Overall, the differences between the PINLED and RN models remain small in the shadow observables, but more pronounced for small masses, while increasing with charge.

\section{Light Deflection Angles}
\label{lightdefangle}
In this section we  will turn to unbound trajectories and analyze light deflection resulting by PINLED $Y^n$ BHs. The trajectories of a photon moving in a static, spherically symmetric spacetime can be conveniently analyzed using orbital equation \eqref{MechEq-Geom} with the effective potential \eqref{WeffLightRays} 
\begin{eqnarray}
   \left( \frac{d \rho}{d \phi} \right)^2  +\rho^2 f(\rho) - \frac{\rho^4}{b^2} = 0
   \label{orbiteq}
\end{eqnarray}
where $b=L/E$. The turning point of the trajectory corresponds to the point of closest approach $\rho_m$, where the radial component of the photon’s four-velocity vanishes. Thus,  by applying Eq. \eqref{orbiteq} to $\rho=\rho_m$  we obtain the following relation among $\rho_m$ and $b$:
\begin{equation}
    \left( \frac{d \rho}{d \phi} \right)^2 |_{\rho=\rho_m} =0 \quad \Rightarrow \quad \frac{1}{b^2}=\frac{f(\rho_m)}{\rho_m^2}.
\end{equation}
Therefore, we can use this relation to calculate first the  asymptotic direction of a light ray with respect to the direction of its closest approach, which we are free to choose as $\phi=0$:
\begin{eqnarray}
    \Delta \phi=\int_{\rho_m}^\infty \frac{d\rho}{ \sqrt{\frac{\rho^4}{\rho_m^2} f(\rho_m)-\rho^2f(\rho)} }
    \label{AsymptoticAngle}
\end{eqnarray}
for a trajectory symmetric about the point of closest approach, the deflection angle $\delta$ of a light ray with respect to its asymptotic incoming direction, can be written as 
\begin{equation}
  \delta = 2\Delta \phi - \pi 
 \label{defa}
 \end{equation}
 As always, the integral for  $\Delta \phi$  in  \eqref{defa} should be calculated for deflection by $Y^n$ BHs after the change of variables $\rho(y)$. The deflection angle is specific to the light trajectory characterized by $\rho_m$, or alternatively by the impact parameter $b$.  Fig.~(\ref{fig:defangle}) illustrates the variation of the deflection angle $\delta$ (in radians) as a function of the closest approach distance $\rho_m$ for light deflection by the $Y^n$ BHs, considering different nonlinearity orders $n=2,3$ and $4$. The results are compared with the Schwarzschild and RN cases, shown, respectively, by the dashed and solid black curves, for a fixed black hole mass $m_{BH}=3$.

As expected, the deflection angle increases sharply as $\rho_m$ approaches the photon sphere radius $\rho_{ps}$, where the light rays winds around the BH with increasingly large number of times. Since the photon sphere radii depend on the BH parameters, each curve has a different divergence point.  For larger and increasing $\rho_m$, the deflection angle decreases monotonically, and all curves converge to the weak deflection regime where $\delta \rightarrow 0$ asymptotically. Increasing the charge $q$ leads to a stronger deviation from the Schwarzschild curve, reducing the deflection angle at a given $\rho_m$. This reflects the repulsive PINLED contribution to the effective gravitational pull, which effectively weakens the light deflection by $Y^n$ BHs.
 However, a comparison with the RN case shows that the deflection angle remains essentially unchanged at high charge values, while deviations becoming appreciable for sufficiently small charges. We also note that the difference between the two curves of the deflection angle  increases while approaching from large distances to the photon sphere. This is also consistent with the observation that the departure of $Y^n$ metric tensor from the RN one, also increses towards the center.

\begin{figure}[h]
\centering
\begin{subfigure}{0.33\textwidth}
    \includegraphics[width=\linewidth, height=\linewidth]{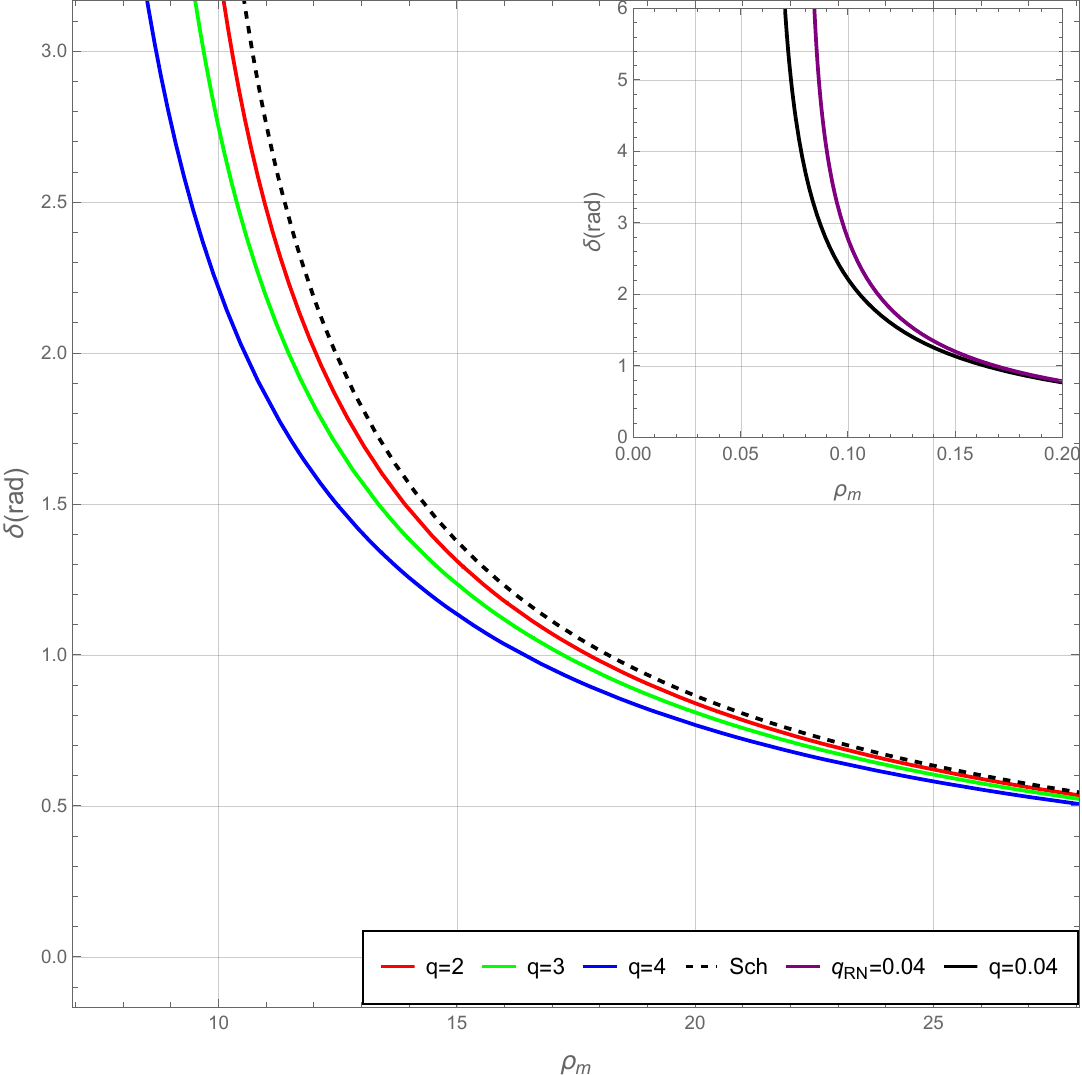} 
    \caption{$n=2$}
    \label{fig:defn2}
\end{subfigure}%
\begin{subfigure}{0.33\textwidth}
    \includegraphics[width=\linewidth, height=\linewidth]{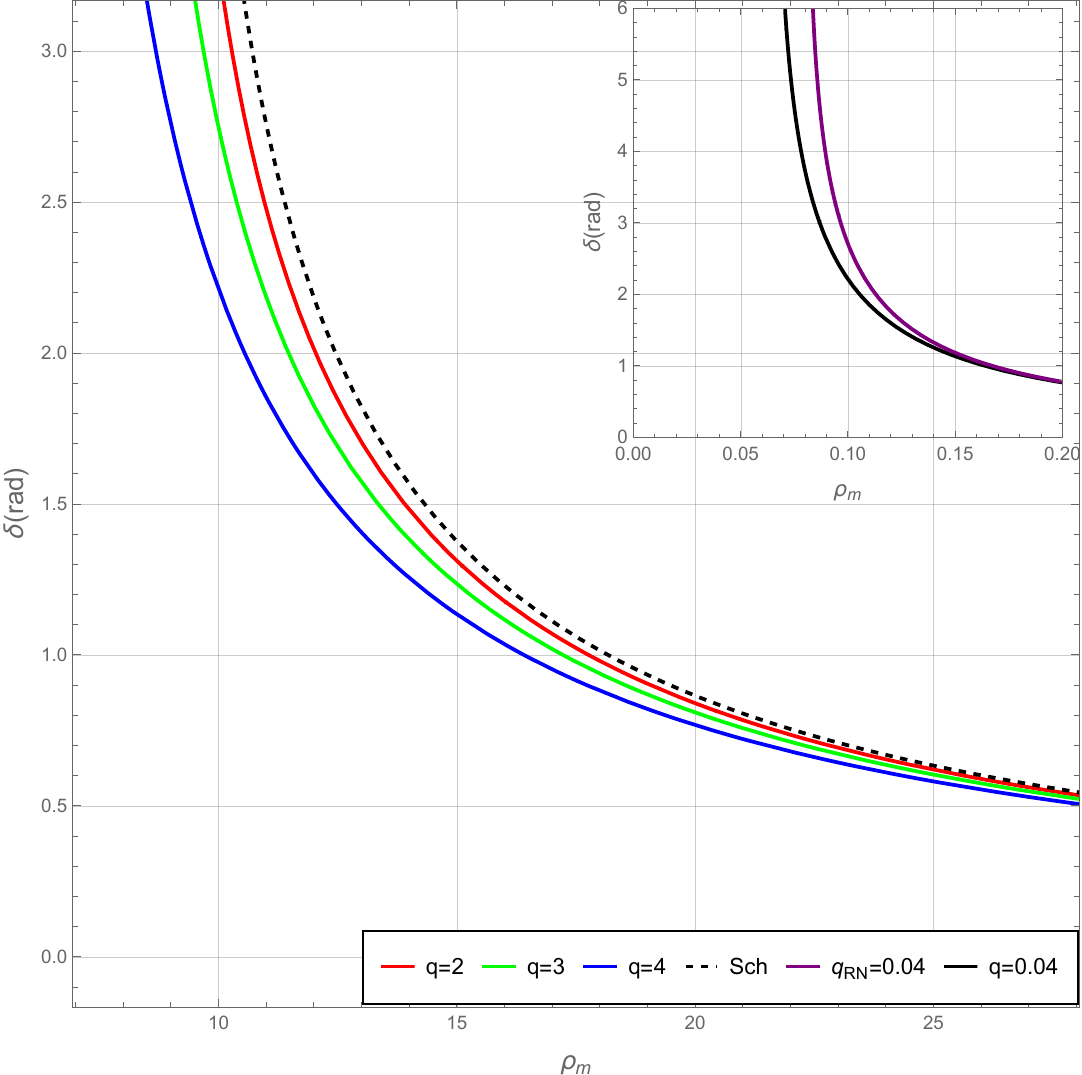}
    \caption{$n=3$}
    \label{fig:defn3}
\end{subfigure}%
\begin{subfigure}{0.33\textwidth}
    \includegraphics[width=\linewidth, height=\linewidth]{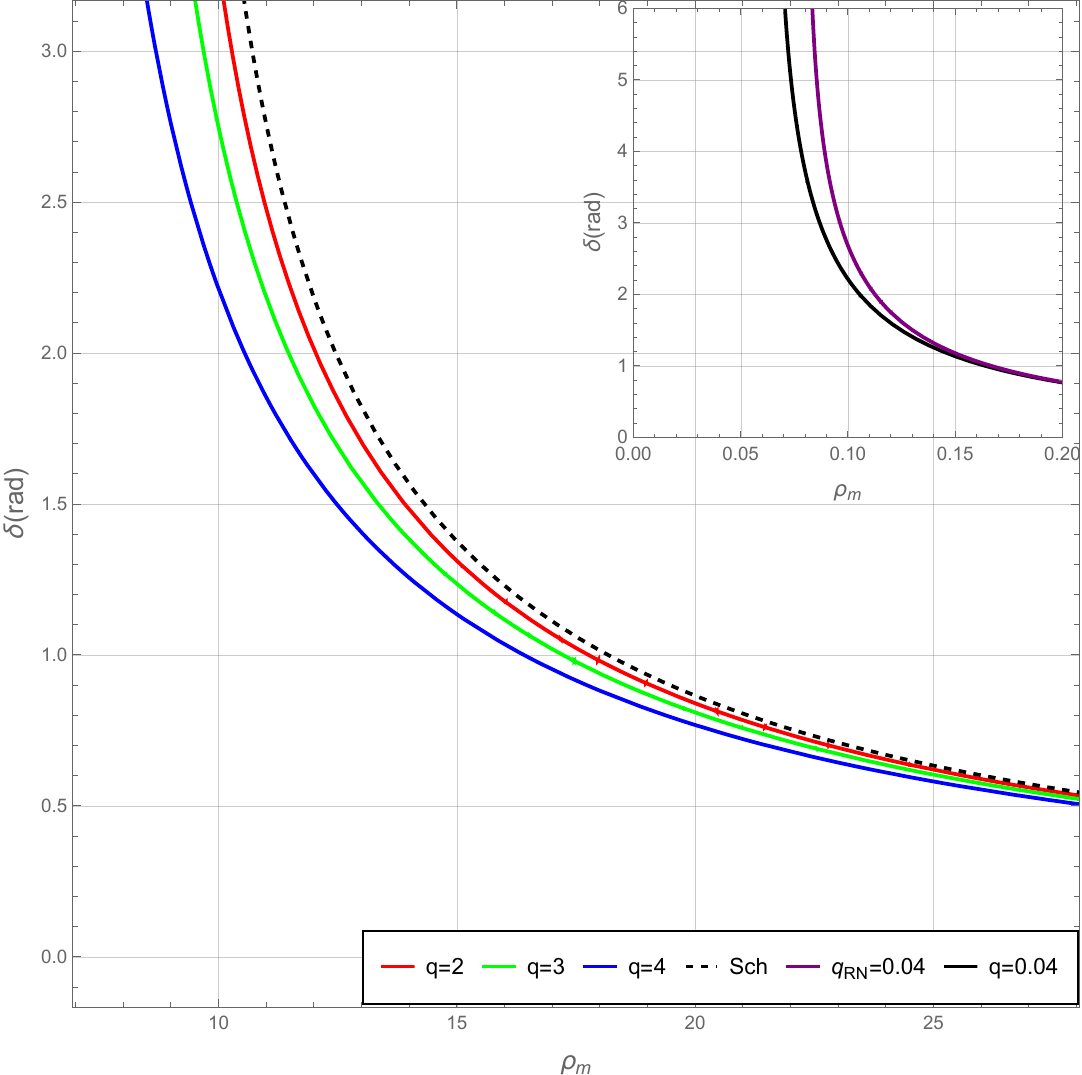} 
    \caption{$n=4$}
    \label{fig:defn4}
\end{subfigure}

\caption{Light deflection angle $\delta$ (in radians) vs closest approach $\rho_m$ for $m_{BH}=3$ and various $q$ values.  Dashed black lines represent Schwarzchild ($q=0$) case. The inserts present a comparison between RN (black) and $Y^n$ (purple) cases, both with small charge and mass values where the difference is visible: $q=0.04$ , $m_{BH}=0.03$.}
\label{fig:defangle}
\end{figure}

\section{periastron Precession}
\label{periastron}

        In this section, we present a general and systematic derivation of the periastron precession within the PINLED model. We focus on the motion of massive test particles, corresponding to the substitution $\xi=1$ in Eqs \eqref{VeffMassive} and \eqref{Weff-Geom}. 
         Eqs. \eqref{MechEq-Geom} and \eqref{Weff-Geom} consist together the  first order ``mechanical'' equation which determines the geometrical shape of the orbits $\rho (\phi)$. The equation for massive test particles is: 
\begin{eqnarray}
   \left(\frac{d\rho}{d\phi}\right)^2=
   -\rho^2 f(\rho) - \frac{\rho^4}{L^2} f(\rho) +\frac{E^2}{L^2}\rho^4
\equiv -W(\rho)\,,
\label{-ww} 
\end{eqnarray}
where we denote here by $W(\rho)$ the effective potential with $\xi=1$. For studying the periastron precession we focus on bound (but not necessarily closed) orbits. The two radial turning points of these orbits are determined by the condition $W(\rho)=0$. These roots define the extrema of the radial motion, with $\rho_-$ corresponding to the periastron and $\rho_+$ to the apastron.  

The angular displacement accumulated as the particle moves from the periastron to the apastron is obtained by integrating $d\phi/d\rho$ using the explicit Eq.\eqref{-ww},  between the corresponding turning points. The total angular advance over one complete radial oscillation (periastron to apastron and back) is therefore given by
\begin{eqnarray}
\Phi
=
2\int_{\rho_-}^{\rho_+}
\frac{d\rho}{\sqrt{-W(\rho)}}\,,
\end{eqnarray}
 For a closed (periodic) Keplerian orbit one would have $\Phi=2\pi$; any deviation from this value signals the presence of periastron precession. Accordingly, the periastron shift per orbit is defined as
\begin{eqnarray}
  \delta\phi
=
\Phi - 2\pi
=
2\int_{\rho_-}^{\rho_+}
\frac{d\rho}{\sqrt{-W(\rho)}} - 2\pi\,. \label{delhpii}
\end{eqnarray}
Notice that the integration limits $\rho_-$ and $\rho_+$ are determined by the orbit parameters: energy $E$ and angular momentum $L$. This is done from the two solutions of the equation $W(\rho_\pm)=0$ mentioned above. However, it is easier to do the converse and solve the same equations for $E$ and $L$  in terms of $\rho_-$ and $\rho_+$. The result is:
\begin{eqnarray}
    E^2 = \frac{f(\rho_-) f(\rho_+)\left(\rho_-^2-\rho_+^2\right)}
    {f(\rho_+)\rho_-^2-f(\rho_-)\rho_+^2}\,,
    \;\;\;
    L^2 = \frac{\left[f(\rho_-)-f(\rho_+)\right]\rho_-^2\rho_+^2}
    {f(\rho_+)\rho_-^2-f(\rho_-)\rho_+^2}\,,
    \label{eandl}
\end{eqnarray}
 For a given orbital configuration specified by the turning points $\rho_-$ and $\rho_+$, these expressions allow the conserved quantities to be determined uniquely. Substituting Eq.~\eqref{eandl} into Eq.~\eqref{delhpii}, one can compute the periastron precession for any bound orbit in any static, spherically symmetric spacetime characterized by a lapse function $f(\rho)$. In the present context of PINLED $Y^n$ BHs, we have to adapt these results for the parametric representation that we are using. This is done in a straightforward way by the change of variables $\rho(y)$ in the integration in Eq.~\eqref{delhpii} together with changing the integration limits from $\rho_-$ and $\rho_+$ to their corresponding values $y_-$ and $y_+$.

\begin{figure}[h!]
\centering
\begin{subfigure}{0.33\textwidth}
    \includegraphics[width=\linewidth, height=\linewidth]{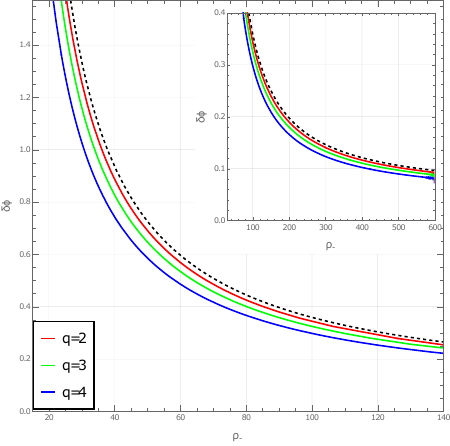} 
    \caption{$\delta\phi$ vs $\rho_-$ for $n=2$ with $\rho_+=600$ and $m_{BH}=3$.}
    \label{fig:precq2}
\end{subfigure}%
\begin{subfigure}{0.33\textwidth}
    \includegraphics[width=\linewidth, height=\linewidth]{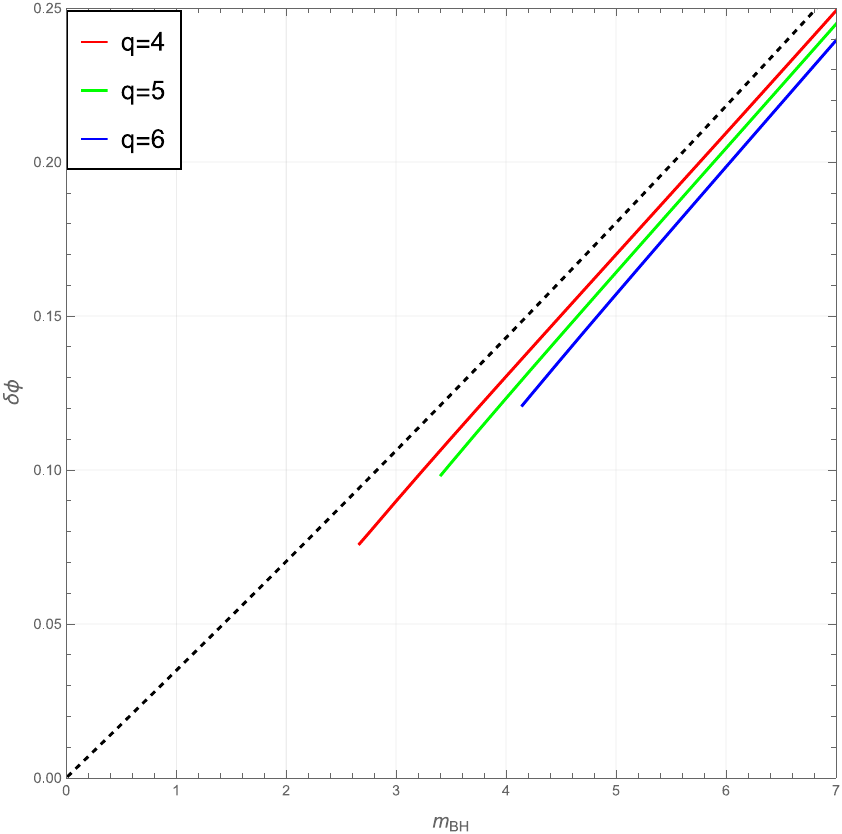}
    \caption{$\delta\phi$ vs $m_{BH}$ for $n=2$ with $\rho_+=600$ and $\rho_-=500$.}
    \label{fig:precq3}
\end{subfigure}%
\begin{subfigure}{0.33\textwidth}
    \includegraphics[width=\linewidth, height=\linewidth]{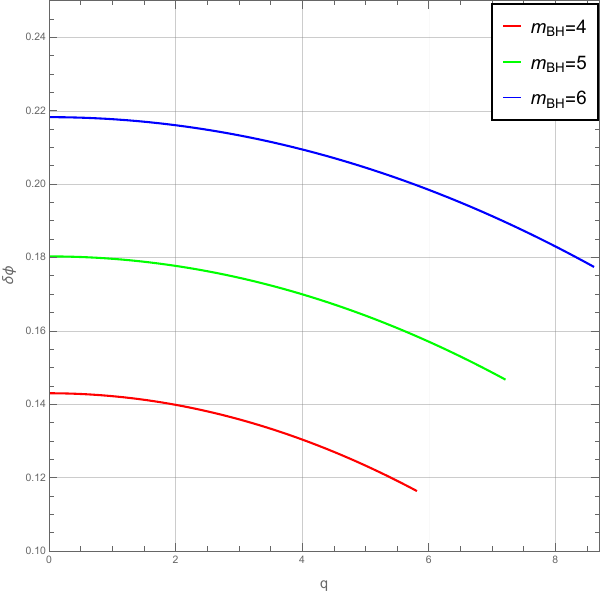} 
    \caption{$\delta\phi$ vs $q$ for $n=2$ with $\rho_{+}=600$ and $\rho_{-}=500$.}
    \label{fig:precq4}
\end{subfigure}

\caption{Periastron precession $\delta\phi$ as a function of the periastron radius $\rho_-$, mass $m_{BH}$ and the charge parameter $q$ for $Y^n$ BHs for fixed apastron values. The black dashed curves  in Fig.~\ref{fig:precq2} and \ref{fig:precq3} correspond to the Schwarzchild BHs, and  the  $q=0$  values of the curves on the vertical axis in  Fig.~\ref{fig:precq4} represent the Schwarzchild precession values with the corresponding $m_{BH}$. }
\label{fig:phip}
\end{figure}
\begin{table}[b!]
\centering
\begin{tabular}{|c|ccc|ccc|}
\hline
 & \multicolumn{3}{c|}{RN} & \multicolumn{3}{c|}{PINLED ($n=2$)} \\
\cline{2-7}
$q$ & $m_{BH}=3$ & $m_{BH}=4$ & $m_{BH}=5$ 
    & $m_{BH}=3$ & $m_{BH}=4$ & $m_{BH}=5$ \\
\hline
1 & 0.10528 & 0.14218 & 0.17961 & 0.10528 & 0.14217 & 0.17960 \\
3 & 0.09707 & 0.13588 & 0.17445 & 0.09705 & 0.13587 & 0.17445 \\
4 & 0.08987 & 0.13036 & 0.16994 & 0.08986 & 0.13034 & 0.16994 \\
\hline
\hline
 & \multicolumn{3}{c|}{PINLED ($n=3$)} & \multicolumn{3}{c|}{PINLED ($n=4$)} \\
\cline{2-7}
$q$ & $m_{BH}=3$ & $m_{BH}=4$ & $m_{BH}=5$
    & $m_{BH}=3$ & $m_{BH}=4$ & $m_{BH}=5$ \\
\hline
1 & 0.10524 & 0.14214 & 0.17957 & 0.10523 & 0.14213 & 0.17956 \\
3 & 0.09677 & 0.13588 & 0.17445 & 0.09667 & 0.13545 & 0.17401 \\
4 & 0.08938 & 0.12981 & 0.16936 & 0.08922 & 0.12963 & 0.16917 \\
\hline
\end{tabular}
\caption{Periastron precession $\delta \phi$ for various values of $n$, $q$ and $m_{BH}$ for $\rho_{-}=500$ and $\rho_{+}=600$.}
\label{table1}
\end{table}

Fig.~\ref{fig:phip} displays the periastron precession $\delta \phi$ for $Y^n$ black holes as a function of the periastron radius $\rho_-$, black hole mass $m_{BH}$ and charge $q$ with fixed apastron $\rho_+.$ 
In Fig.~\ref{fig:precq2} $\delta \phi$ decreases monotonically with respect to $\rho_-$ for fixed $m_{BH}$ and several values of $q$.  Increasing $q$ yields smaller precession for fixed $\rho_-$.  We notice that at the end points of the curves which correspond to circular orbits, $\delta \phi$ does not vanish. This is a well-known peculiarity of the precession angle. Fig.~\ref{fig:precq3} shows that $\delta \phi$  grows with $m_{BH}$ similarly to the Schwarzschild case, nearly linear in $m_{BH}$ for fixed $q$. The three curves for different $q$ values remain closely spaced, so demonstrating that the charge contribution is a subdominant correction to the mass term, since for fixed $q$, the mass can grow indefinitely and the metric becomes closer to Schwarzchild. On the other hand, the charge value has a bound of the order of $m_{BH}$. Fig.~\ref{fig:precq4} illustrates the dependence on $q$ at fixed $m_{BH}$ and $\rho_-$ where the precession decreases gradually with increasing $q$. This is qualitatively in line with the RN result, where the charge introduces a retrograde correction to the precession angle. For the higher $n$-values,  $n=3$ and $n=4$, the precession results are indistinguishable from the $n=2$ case within the resolution of the plot. Table~\ref{table1} compares the periastron precession between the RN and PINLED models for the same $\rho_+$ and $\rho_-$ across various values of $n,q$ and $m_{BH}$. One can see a systematic deviation in PINLED values that grows with both $n$ and $q$. In all cases $\delta \phi$ increases with $m_{BH}$ and decreases with $q$.

\section{Conclusion}
\label{conclusion}

\begin{table}[b!]
\centering
\begin{tabular}{|c|c|c|c|c|c|c|c|c|c|c|c|c|}
\hline
$q$ & \multicolumn{4}{c|}{$n=2$} & \multicolumn{4}{c|}{$n=3$} & \multicolumn{4}{c|}{$n=4$} \\
\cline{2-13}
 & $\rho_{H}$ & $\rho_{\rm ISCO}$ & $\rho_{ps}$ & $\rho_s$  & $\rho_{H}$ & $\rho_{\rm ISCO}$ & $\rho_{ps}$ & $\rho_s$ & $\rho_{H}$ & $\rho_{ISCO}$ & $\rho_{ps}$ & $\rho_s$ \\
\hline
1 & 5.9155 & 17.7475 & 8.8875 & 15.4425 & 5.91548 & 17.7475 & 8.88748 & 15.4425 & 5.91548 & 17.7475& 8.88748 & 15.4425 \\
3 & 5.12662 & 15.5007& 7.85661 & 14.1304 & 5.12137 & 15.4993& 7.85411 & 14.12880 & 5.12132 & 15.4993 & 7.8541 & 14.1288\\
3.55021 & 4.66136 & 14.2943 & 7.27317 & 13.4144 & 4.64298 & 14.2903& 7.26519 & 13.4095 & 4.64258 & 14.2903& 7.26515 & 13.4096 \\
3.6 & 4.60883 & 14.1652 & 7.20899 & 13.3372 & 4.58796 & 14.1608& 7.20005 & 13.3318 & 4.58747 & 14.1608 & 7.2 & 13.3318 \\
3.75 & 4.4355 & 13.7219 & 7.00031 & 13.0887 & 4.40414 & 13.7158& 6.98756 & 13.0811 & 4.40316 & 13.7157& 6.98747 & 13.0811 \\
4 & 4.07624 & 12.9592 & 6.58674 & 12.6095 & 4.00442 & 12.9484& 6.56181 & 12.5953 & 4.0003 & 12.9484 & 6.56156 & 13.0811 \\
\hline
\end{tabular}
\caption{Characteristic radii $\rho_H$, $\rho_{\rm ISCO}$, $\rho_{\rm ps}$, and $\rho_s$ for different values of the charge parameter $q$ and nonlinearity index $n=2,3,4$, computed for a fixed black hole mass $m_{\rm BH}=3$.}
\label{table}
\end{table}

In this work our main aim was to provide a systematic and self-contained set of strong and weak field observables for the static, spherically symmetric PINLED--$Y^n$ black-hole solutions minimally coupled to Einstein--Hilbert gravity. Motivated by the EHT era, where photon-ring/shadow measurements and precision lensing tests increasingly probe the near-horizon geometry, we analyzed null and timelike geodesics in the spacetime \eqref{line-element} (with the metric function given by parametric representation \eqref{BHmetricYn} with \eqref{BHmassFunctionYn}+\eqref{BHfieldEnergyYn}), and quantified how the charge parameter $q$ and the nonlinearity index $n$ deform the photon sphere, the shadow size, and the characteristic radii governing massive particle motion like the ISCO and perihelion precession angle.

For timelike geodesics, we determined circular orbits and the ISCO from $V'_{\rm eff}=0$ and $V''_{\rm eff}=0$, obtaining the compact condition \eqref{ISCOEqFinal} and solving it 
for $\rho_{\rm ISCO}$ as a function of mass, charge and the index $n$ as showed in subsection \ref{massiveneutral}. Table~\ref{table} demonstrates that $\rho_{\rm ISCO}$ decreases with increasing $q$ over the parameter range displayed, tracking the contraction of the characteristic strong field region and providing an additional, independent probe of the geometry. Since the ISCO sets a key scale for accretion dynamics (e.g., the inner edge of thin disks and the onset of strong-field orbital shear), these shifts translate into potentially observable changes in disk spectra and variability when PINLED corrections are appreciable. In short, both in null and timelike geodesics, we see the inherent electromagnetic repulsive effect.

Our results show that, for fixed black hole mass, increasing the charge parameter $q$ shifts the characteristic radii inward: the horizon radius $\rho_H$, the photon sphere radius $\rho_{\rm ps}$, the shadow radius $\rho_s$, and the ISCO radius $\rho_{\rm ISCO}$ all decrease as $q$ grows. This behavior is consistent with the fact that the nonlinear electromagnetic sector reduces the net attractive pull relative to the Schwarzschild case. Conversely, for fixed $q$, these radii increase with $m_{\rm BH}$, as expected from the stronger gravitational field of a more massive source.

Another outcome of our analysis is that null-geodesic observables showed in subsection~\ref{massless} and~\ref{shadow} provide a probe of departures from the Reissner-Nordstr\"om geometry.  provide the clearest probe of departures from the Reissner-Nordstr\"om geometry. In particular, both $\rho_{\rm ps}$ and $\rho_s$ decrease monotonically with $q$, while the differences between the PINLED and RN cases become most visible in the low mass high charge regime. By contrast, the ISCO remains very close to the RN prediction over most of the parameter space explored here, especially for $n=3$ and $4$, indicating that timelike circular motion is a comparatively weak discriminator of the nonlinear PINLED corrections. The dependence on the nonlinearity index $n$ is generally subleading compared with the dependence on $q$. Although increasing $n$ produces small quantitative shifts in the characteristic radii, these corrections remain modest, and for several observables the $n=3$ and $n=4$ cases become nearly indistinguishable from the RN limit. This indicates that, within the present family of solutions, the charge parameter is the dominant control parameter for the optical and orbital structure, while the effect of the nonlinear index is comparatively mild.

We also analyzed light deflection in section \ref{lightdefangle} and periastron precesion in section \ref{periastron}. As usual, the light deflection  weakens as the light ray closest approach increases. For fixed closest-approach distance, fixed BH mass and increasing charge, the light deflection angle decreases from the maximum of the Schwarzschild value in the zero-charge limit, while its dependence on $n$ remains weak. The deviation of light deflection for $Y^n$ BHs with respect to RN is minute in the weak field - large $\rho_m$ values. It becomes distinguishable in the opposite limit where $\rho_m$ gets closer to the photon sphere and especially for small BH masses. Likewise, for bound particle orbits with fixed turning points, the periastron advance increases with $m_{\rm BH}$ but decreases with $q$, again with only small corrections induced by changing $n$. These particle observables, ISCO and especially the periastron precession provide useful consistency checks, but they are less sensitive to the PINLED deformation than the quantities associated with the photon region.

Overall, our results furnish a coherent set of strong and weak field calculations for this new class of first-order nonlinear-electrodynamics black holes. They show that the most promising signatures of the PINLED-$Y^n$ geometry are encoded in null-geodesic observables, especially the photon sphere and the shadow, whereas timelike circular motion and weak field tests mainly provide secondary constraints. Natural extensions of this work include rotating PINLED solutions, quasinormal-mode and ringdown analyses, and the incorporation of realistic emission models and radiative-transfer effects for direct comparison with horizon-scale observations.

\acknowledgments
{\.I}.{\.I}.{\c C}. and  Y.V. would like to thank for financial support by the Israeli fund for scientific regional cooperation and the Research Authority of the Open University of Israel. A. \"O. would like to acknowledge networking support of the COST Action CA21106 - COSMIC WISPers in the Dark Universe: Theory, astrophysics and experiments (CosmicWISPers), the COST Action CA22113 - Fundamental challenges in theoretical physics (THEORY-CHALLENGES), the COST Action CA21136 - Addressing observational tensions in cosmology with systematics and fundamental physics (CosmoVerse), the COST Action CA23130 - Bridging high and low energies in search of quantum gravity (BridgeQG), and the COST Action CA23115 - Relativistic Quantum Information (RQI) funded by COST (European Cooperation in Science and Technology). A. \"O. also thanks to EMU, TUBITAK, ULAKBIM (Turkiye) and SCOAP3 (Switzerland) for their support. Y.V. also thanks COST Action CA23130 - Bridging high and low energies in search of quantum gravity (BridgeQG).

\bibliography{references}

@article{Plebanski1970,
    author = {Plebanski, J.},
    title = "{Lectures on Non-Linear Electrodynamics}",
    journal = "Nordita, Copenhagen",
    year = "1970"
}

@article{Rasanen-Verbin2022,
    author ={Rasanen, S. and Verbin, Y.},
    title = "{Palatini formulation for gauge theory: implications for slow-roll inflation}",
    eprint = "2211.15584",
    archivePrefix = "arXiv",
    primaryClass = "astro-ph.CO",
    doi = "10.21105/astro.2211.15584",
    journal = "Open Journal of Astrophysics",
    volume = "6",
    year = "2023"
}

@article{EHTM87I2019,
    author = "Akiyama, Kazunori and others",
    collaboration = "Event Horizon Telescope",
    title = "{First M87 Event Horizon Telescope Results. I. The Shadow of the Supermassive Black Hole}",
    eprint = "1906.11238",
    archivePrefix = "arXiv",
    primaryClass = "astro-ph.GA",
    doi = "10.3847/2041-8213/ab0ec7",
    journal = "Astrophys. J. Lett.",
    volume = "875",
    pages = "L1",
    year = "2019"
}

@article{EHTSgrAI2022,
    author = "Akiyama, Kazunori and others",
    collaboration = "Event Horizon Telescope",
    title = "{First Sagittarius A* Event Horizon Telescope Results. I. The Shadow of the Supermassive Black Hole in the Center of the Milky Way}",
    eprint = "2311.08680",
    archivePrefix = "arXiv",
    primaryClass = "astro-ph.HE",
    doi = "10.3847/2041-8213/ac6674",
    journal = "Astrophys. J. Lett.",
    volume = "930",
    number = "2",
    pages = "L12",
    year = "2022"
}

@article{Synge1966,
    author = "Synge, J. L.",
    title = "{The Escape of Photons from Gravitationally Intense Stars}",
    doi = "10.1093/mnras/131.3.463",
    journal = "Mon. Not. Roy. Astron. Soc.",
    volume = "131",
    number = "3",
    pages = "463--466",
    year = "1966"
}

@article{Virbhadra:1999nm,
    author = "Virbhadra, K. S. and Ellis, George F. R.",
    title = "{Schwarzschild black hole lensing}",
    eprint = "astro-ph/9904193",
    archivePrefix = "arXiv",
    doi = "10.1103/PhysRevD.62.084003",
    journal = "Phys. Rev. D",
    volume = "62",
    pages = "084003",
    year = "2000"
}

@article{Lambiase:2024lvo,
    author = {Lambiase, Gaetano and Gogoi, Dhruba Jyoti and Pantig, Reggie C. and {\"O}vg{\"u}n, Ali},
    title = "{Shadow and quasinormal modes of the rotating Einstein{\textendash}Euler{\textendash}Heisenberg black holes}",
    eprint = "2406.18300",
    archivePrefix = "arXiv",
    primaryClass = "gr-qc",
    doi = "10.1016/j.dark.2025.101886",
    journal = "Phys. Dark Univ.",
    volume = "48",
    pages = "101886",
    year = "2025"
}

@article{Uniyal:2023inx,
    author = {Uniyal, Akhil and Chakrabarti, Sayan and Pantig, Reggie C. and {\"O}vg{\"u}n, Ali},
    title = "{Nonlinearly charged black holes: Shadow and thin-accretion disk}",
    eprint = "2303.07174",
    archivePrefix = "arXiv",
    primaryClass = "gr-qc",
    doi = "10.1016/j.newast.2024.102249",
    journal = "New Astron.",
    volume = "111",
    pages = "102249",
    year = "2024"
}

@article{Boyeneni:2025tsx,
    author = "Boyeneni, Siddharth and Wu, Jiaxi and Most, Elias R.",
    title = "{Unveiling the Electrodynamic Nature of Spacetime Collisions}",
    eprint = "2504.15978",
    archivePrefix = "arXiv",
    primaryClass = "gr-qc",
    doi = "10.1103/995s-wxl7",
    journal = "Phys. Rev. Lett.",
    volume = "135",
    number = "10",
    pages = "101401",
    year = "2025"
}

@article{Claudel:2000yi,
    author = "Claudel, Clarissa-Marie and Virbhadra, K. S. and Ellis, G. F. R.",
    title = "{The Geometry of photon surfaces}",
    eprint = "gr-qc/0005050",
    archivePrefix = "arXiv",
    doi = "10.1063/1.1308507",
    journal = "J. Math. Phys.",
    volume = "42",
    pages = "818--838",
    year = "2001"
}

@article{Capozziello:2024ucm,
    author = "Capozziello, Salvatore and De Bianchi, Silvia and Battista, Emmanuele",
    title = "{Avoiding singularities in Lorentzian-Euclidean black holes: The role of~atemporality}",
    eprint = "2404.17267",
    archivePrefix = "arXiv",
    primaryClass = "gr-qc",
    doi = "10.1103/PhysRevD.109.104060",
    journal = "Phys. Rev. D",
    volume = "109",
    number = "10",
    pages = "104060",
    year = "2024"
}

@article{Capozziello:2025wwl,
    author = "Capozziello, Salvatore and Battista, Emmanuele and De Bianchi, Silvia",
    title = "{Null geodesics, causal structure, and matter accretion in Lorentzian-Euclidean black holes}",
    eprint = "2507.08431",
    archivePrefix = "arXiv",
    primaryClass = "gr-qc",
    doi = "10.1103/ybjp-8w2w",
    journal = "Phys. Rev. D",
    volume = "112",
    number = "4",
    pages = "044009",
    year = "2025"
}

@article{Virbhadra:2008ws,
    author = "Virbhadra, K. S.",
    title = "{Relativistic images of Schwarzschild black hole lensing}",
    eprint = "0810.2109",
    archivePrefix = "arXiv",
    primaryClass = "gr-qc",
    doi = "10.1103/PhysRevD.79.083004",
    journal = "Phys. Rev. D",
    volume = "79",
    pages = "083004",
    year = "2009"
}

@article{Gravity2020S2,
    author = "Abuter, R. and others",
    collaboration = "GRAVITY",
    title = "{Detection of the Schwarzschild precession in the orbit of the star S2 near the Galactic centre massive black hole}",
    eprint = "2004.07187",
    archivePrefix = "arXiv",
    primaryClass = "astro-ph.GA",
    doi = "10.1051/0004-6361/202037813",
    journal = "Astron. Astrophys.",
    volume = "636",
    pages = "L5",
    year = "2020"
}

@article{Bozza2002StrongField,
  author = "Bozza, V.",
    title = "{Gravitational lensing in the strong field limit}",
    eprint = "gr-qc/0208075",
    archivePrefix = "arXiv",
    doi = "10.1103/PhysRevD.66.103001",
    journal = "Phys. Rev. D",
    volume = "66",
    pages = "103001",
    year = "2002"
}

@article{CunhaHerdeiro:2018,
  author = "Cunha, Pedro V. P. and Herdeiro, Carlos A. R.",
    title = "{Shadows and strong gravitational lensing: a brief review}",
    eprint = "1801.00860",
    archivePrefix = "arXiv",
    primaryClass = "gr-qc",
    doi = "10.1007/s10714-018-2361-9",
    journal = "Gen. Rel. Grav.",
    volume = "50",
    number = "4",
    pages = "42",
    year = "2018"
}

@article{PerlickTsupko:2022,
   author = "Perlick, Volker and Tsupko, Oleg Yu.",
    title = "{Calculating black hole shadows: Review of analytical studies}",
    eprint = "2105.07101",
    archivePrefix = "arXiv",
    primaryClass = "gr-qc",
    doi = "10.1016/j.physrep.2021.10.004",
    journal = "Phys. Rept.",
    volume = "947",
    pages = "1--39",
    year = "2022"
}

@article{Toshmatov:2021fgm,
    author = "Toshmatov, Bobir and Ahmedov, Bobomurat and Malafarina, Daniele",
    title = "{Can a light ray distinguish charge of a black hole in nonlinear electrodynamics?}",
    eprint = "2101.05496",
    archivePrefix = "arXiv",
    primaryClass = "gr-qc",
    doi = "10.1103/PhysRevD.103.024026",
    journal = "Phys. Rev. D",
    volume = "103",
    number = "2",
    pages = "024026",
    year = "2021"
}

@article{Vagnozzi:2023,
  author = "Vagnozzi, Sunny and others",
    title = "{Horizon-scale tests of gravity theories and fundamental physics from the Event Horizon Telescope image of Sagittarius A}",
    eprint = "2205.07787",
    archivePrefix = "arXiv",
    primaryClass = "gr-qc",
    reportNumber = "UCI-HEP-TR-2022-07",
    doi = "10.1088/1361-6382/acd97b",
    journal = "Class. Quant. Grav.",
    volume = "40",
    number = "16",
    pages = "165007",
    year = "2023"
}

@article{Verbin:2025PNLED,
  author = {Verbin, Yosef and Pulice, Beyhan and {\"O}vg{\"u}n, Ali and Huang, Hyat},
    title = "{New black hole solutions of second and first order formulations of nonlinear electrodynamics}",
    eprint = "2412.20989",
    archivePrefix = "arXiv",
    primaryClass = "gr-qc",
    doi = "10.1103/PhysRevD.111.084061",
    journal = "Phys. Rev. D",
    volume = "111",
    number = "8",
    pages = "084061",
    year = "2025"
}

@article{AyonBeatoGarcia:1998,
   author = "Harada, Tomohiro",
    title = "{Neutron stars in scalar tensor theories of gravity and catastrophe theory}",
    eprint = "gr-qc/9801049",
    archivePrefix = "arXiv",
    reportNumber = "KUNS-1476",
    doi = "10.1103/PhysRevD.57.4802",
    journal = "Phys. Rev. D",
    volume = "57",
    pages = "4802--4811",
    year = "1998"
}

@article{Euler-Kockel,
   author = "H.~Euler and B.~Kockel",
    title = "{The scattering of light by light in Dirac's theory}",
    journal = "Naturwiss",
    volume = "23",
    pages = "246",
    year = "1933"
}

@article{HeisenbergEuler:1936,
   author = "Heisenberg, W. and Euler, H.",
    title = "{Consequences of Dirac's theory of positrons}",
    eprint = "physics/0605038",
    archivePrefix = "arXiv",
    doi = "10.1007/BF01343663",
    journal = "Z. Phys.",
    volume = "98",
    number = "11-12",
    pages = "714--732",
    year = "1936"
}

@article{BornInfeld:1934,
    author = "Born, M. and Infeld, L.",
    title = "{Foundations of the new field theory}",
    doi = "10.1098/rspa.1934.0059",
    journal = "Proc. Roy. Soc. Lond. A",
    volume = "144",
    number = "852",
    pages = "425--451",
    year = "1934"
}

@article{Ayon-Beato:1998hmi,
    author = "Ayon-Beato, Eloy and Garcia, Alberto",
    title = "{Regular black hole in general relativity coupled to nonlinear electrodynamics}",
    eprint = "gr-qc/9911046",
    archivePrefix = "arXiv",
    doi = "10.1103/PhysRevLett.80.5056",
    journal = "Phys. Rev. Lett.",
    volume = "80",
    pages = "5056--5059",
    year = "1998"
}

@article{Ayon-Beato:2000mjt,
    author = "Ayon-Beato, Eloy and Garcia, Alberto",
    title = "{The Bardeen model as a nonlinear magnetic monopole}",
    eprint = "gr-qc/0009077",
    archivePrefix = "arXiv",
    doi = "10.1016/S0370-2693(00)01125-4",
    journal = "Phys. Lett. B",
    volume = "493",
    pages = "149--152",
    year = "2000"
}

@article{Bronnikov:2000vy,
    author = "Bronnikov, Kirill A.",
    title = "{Regular magnetic black holes and monopoles from nonlinear electrodynamics}",
    eprint = "gr-qc/0006014",
    archivePrefix = "arXiv",
    doi = "10.1103/PhysRevD.63.044005",
    journal = "Phys. Rev. D",
    volume = "63",
    pages = "044005",
    year = "2001"
}

@article{Ayon-Beato:1999kuh,
    author = "Ayon-Beato, Eloy and Garcia, Alberto",
    title = "{New regular black hole solution from nonlinear electrodynamics}",
    eprint = "hep-th/9911174",
    archivePrefix = "arXiv",
    doi = "10.1016/S0370-2693(99)01038-2",
    journal = "Phys. Lett. B",
    volume = "464",
    pages = "25",
    year = "1999"
}

@article{Gibbons:1995cv,
    author = "Gibbons, G. W. and Rasheed, D. A.",
    title = "{Electric - magnetic duality rotations in nonlinear electrodynamics}",
    eprint = "hep-th/9506035",
    archivePrefix = "arXiv",
    doi = "10.1016/0550-3213(95)00409-L",
    journal = "Nucl. Phys. B",
    volume = "454",
    pages = "185--206",
    year = "1995"
}

@article{Fan:2016hvf,
    author = "Fan, Zhong-Ying and Wang, Xiaobao",
    title = "{Construction of Regular Black Holes in General Relativity}",
    eprint = "1610.02636",
    archivePrefix = "arXiv",
    primaryClass = "gr-qc",
    doi = "10.1103/PhysRevD.94.124027",
    journal = "Phys. Rev. D",
    volume = "94",
    number = "12",
    pages = "124027",
    year = "2016"
}

@article{Dymnikova:2004zc,
    author = "Dymnikova, Irina",
    title = "{Regular electrically charged structures in nonlinear electrodynamics coupled to general relativity}",
    eprint = "gr-qc/0407072",
    archivePrefix = "arXiv",
    doi = "10.1088/0264-9381/21/18/009",
    journal = "Class. Quant. Grav.",
    volume = "21",
    pages = "4417--4429",
    year = "2004"
}

@article{Balart:2014cga,
    author = "Balart, Leonardo and Vagenas, Elias C.",
    title = "{Regular black holes with a nonlinear electrodynamics source}",
    eprint = "1408.0306",
    archivePrefix = "arXiv",
    primaryClass = "gr-qc",
    doi = "10.1103/PhysRevD.90.124045",
    journal = "Phys. Rev. D",
    volume = "90",
    number = "12",
    pages = "124045",
    year = "2014"
}

@article{Allahyari:2019jqz,
    author = "Allahyari, Alireza and Khodadi, Mohsen and Vagnozzi, Sunny and Mota, David F.",
    title = "{Magnetically charged black holes from non-linear electrodynamics and the Event Horizon Telescope}",
    eprint = "1912.08231",
    archivePrefix = "arXiv",
    primaryClass = "gr-qc",
    doi = "10.1088/1475-7516/2020/02/003",
    journal = "JCAP",
    volume = "02",
    pages = "003",
    year = "2020"
}

@article{Novello:1999pg,
    author = "Novello, M. and De Lorenci, V. A. and Salim, J. M. and Klippert, Renato",
    title = "{Geometrical aspects of light propagation in nonlinear electrodynamics}",
    eprint = "gr-qc/9911085",
    archivePrefix = "arXiv",
    reportNumber = "CBPF-NF-050-99",
    doi = "10.1103/PhysRevD.61.045001",
    journal = "Phys. Rev. D",
    volume = "61",
    pages = "045001",
    year = "2000"
}

@article{Okyay:2021nnh,
    author = {Okyay, Mert and {\"O}vg{\"u}n, Ali},
    title = "{Nonlinear electrodynamics effects on the black hole shadow, deflection angle, quasinormal modes and greybody factors}",
    eprint = "2108.07766",
    archivePrefix = "arXiv",
    primaryClass = "gr-qc",
    doi = "10.1088/1475-7516/2022/01/009",
    journal = "JCAP",
    volume = "01",
    number = "01",
    pages = "009",
    year = "2022"
}

@article{Hassaine:2008pw,
    author = "Hassaine, Mokhtar and Martinez, Cristian",
    title = "{Higher-dimensional charged black holes solutions with a nonlinear electrodynamics source}",
    eprint = "0803.2946",
    archivePrefix = "arXiv",
    primaryClass = "hep-th",
    reportNumber = "CECS-PHY-06-25",
    doi = "10.1088/0264-9381/25/19/195023",
    journal = "Class. Quant. Grav.",
    volume = "25",
    pages = "195023",
    year = "2008"
}

@article{Maeda:2008ha,
    author = "Maeda, Hideki and Hassaine, Mokhtar and Martinez, Cristian",
    title = "{Lovelock black holes with a nonlinear Maxwell field}",
    eprint = "0812.2038",
    archivePrefix = "arXiv",
    primaryClass = "gr-qc",
    reportNumber = "CECS-PHY-08-20",
    doi = "10.1103/PhysRevD.79.044012",
    journal = "Phys. Rev. D",
    volume = "79",
    pages = "044012",
    year = "2009"
}

@article{Gibbons:1995ap,
    author = "Gibbons, G W and Rasheed, D A",
    title = "{Sl(2,R) invariance of nonlinear electrodynamics coupled to an axion and a dilaton}",
    eprint = "hep-th/9509141",
    archivePrefix = "arXiv",
    reportNumber = "DAMTP-R-95-48",
    doi = "10.1016/0370-2693(95)01272-9",
    journal = "Phys. Lett. B",
    volume = "365",
    pages = "46--50",
    year = "1996"
}

@article{Berej:2006cc,
    author = "Berej, Waldemar and Matyjasek, Jerzy and Tryniecki, Dariusz and Woronowicz, Mariusz",
    title = "{Regular black holes in quadratic gravity}",
    eprint = "hep-th/0606185",
    archivePrefix = "arXiv",
    doi = "10.1007/s10714-006-0270-9",
    journal = "Gen. Rel. Grav.",
    volume = "38",
    pages = "885--906",
    year = "2006"
}

@article{Ghosh:2014pba,
    author = "Ghosh, Sushant G.",
    title = "{A nonsingular rotating black hole}",
    eprint = "1408.5668",
    archivePrefix = "arXiv",
    primaryClass = "gr-qc",
    doi = "10.1140/epjc/s10052-015-3740-y",
    journal = "Eur. Phys. J. C",
    volume = "75",
    number = "11",
    pages = "532",
    year = "2015"
}

@article{Gibbons:2000xe,
    author = "Gibbons, G. W. and Herdeiro, C. A. R.",
    title = "{Born-Infeld theory and stringy causality}",
    eprint = "hep-th/0008052",
    archivePrefix = "arXiv",
    reportNumber = "DAMTP-2000-71",
    doi = "10.1103/PhysRevD.63.064006",
    journal = "Phys. Rev. D",
    volume = "63",
    pages = "064006",
    year = "2001"
}

@article{Sorokin:2021tge,
    author = "Sorokin, Dmitri P.",
    title = "{Introductory Notes on Non-linear Electrodynamics and its Applications}",
    eprint = "2112.12118",
    archivePrefix = "arXiv",
    primaryClass = "hep-th",
    doi = "10.1002/prop.202200092",
    journal = "Fortsch. Phys.",
    volume = "70",
    number = "7-8",
    pages = "2200092",
    year = "2022"
}

@article{Eiroa:2005ag,
    author = "Eiroa, Ernesto F.",
    title = "{Gravitational lensing by Einstein-Born-Infeld black holes}",
    eprint = "gr-qc/0511065",
    archivePrefix = "arXiv",
    doi = "10.1103/PhysRevD.73.043002",
    journal = "Phys. Rev. D",
    volume = "73",
    pages = "043002",
    year = "2006"
}

@article{Gonzalez:2009nn,
    author = "Gonzalez, Hernan A. and Hassaine, Mokhtar and Martinez, Cristian",
    title = "{Thermodynamics of charged black holes with a nonlinear electrodynamics source}",
    eprint = "0909.1365",
    archivePrefix = "arXiv",
    primaryClass = "hep-th",
    reportNumber = "CECS-PHY-09-08",
    doi = "10.1103/PhysRevD.80.104008",
    journal = "Phys. Rev. D",
    volume = "80",
    pages = "104008",
    year = "2009"
}

@article{Toshmatov:2017zpr,
    author = "Toshmatov, Bobir and Stuchl{\'\i}k, Zden{\v{e}}k and Ahmedov, Bobomurat",
    title = "{Generic rotating regular black holes in general relativity coupled to nonlinear electrodynamics}",
    eprint = "1704.07300",
    archivePrefix = "arXiv",
    primaryClass = "gr-qc",
    doi = "10.1103/PhysRevD.95.084037",
    journal = "Phys. Rev. D",
    volume = "95",
    number = "8",
    pages = "084037",
    year = "2017"
}

@article{Uniyal:2022vdu,
    author = {Uniyal, Akhil and Pantig, Reggie C. and {\"O}vg{\"u}n, Ali},
    title = "{Probing a non-linear electrodynamics black hole with thin accretion disk, shadow, and deflection angle with M87* and Sgr A* from EHT}",
    eprint = "2205.11072",
    archivePrefix = "arXiv",
    primaryClass = "gr-qc",
    doi = "10.1016/j.dark.2023.101178",
    journal = "Phys. Dark Univ.",
    volume = "40",
    pages = "101178",
    year = "2023"
}

@article{DeLorenci:2000yh,
    author = "De Lorenci, V. A. and Klippert, R. and Novello, M. and Salim, J. M.",
    title = "{Light propagation in nonlinear electrodynamics}",
    eprint = "gr-qc/0005049",
    archivePrefix = "arXiv",
    doi = "10.1016/S0370-2693(00)00522-0",
    journal = "Phys. Lett. B",
    volume = "482",
    number = "1-3",
    pages = "134--140",
    year = "2000"
}

@article{Olmo:2011ja,
    author = "Olmo, Gonzalo J. and Rubiera-Garcia, D.",
    title = "{Palatini $f(R)$ Black Holes in Nonlinear Electrodynamics}",
    eprint = "1110.0850",
    archivePrefix = "arXiv",
    primaryClass = "gr-qc",
    doi = "10.1103/PhysRevD.84.124059",
    journal = "Phys. Rev. D",
    volume = "84",
    pages = "124059",
    year = "2011"
}

@article{Dymnikova:2015hka,
    author = "Dymnikova, Irina and Galaktionov, Evgeny",
    title = "{Regular rotating electrically charged black holes and solitons in non-linear electrodynamics minimally coupled to gravity}",
    eprint = "1510.01353",
    archivePrefix = "arXiv",
    primaryClass = "gr-qc",
    doi = "10.1088/0264-9381/32/16/165015",
    journal = "Class. Quant. Grav.",
    volume = "32",
    number = "16",
    pages = "165015",
    year = "2015"
}

@article{Rodrigues:2015ayd,
    author = "Rodrigues, Manuel E. and Junior, Ednaldo L. B. and Marques, Glauber T. and Zanchin, Vilson T.",
    title = "{Regular black holes in $f(R)$ gravity coupled to nonlinear electrodynamics}",
    eprint = "1511.00569",
    archivePrefix = "arXiv",
    primaryClass = "gr-qc",
    doi = "10.1103/PhysRevD.94.024062",
    journal = "Phys. Rev. D",
    volume = "94",
    number = "2",
    pages = "024062",
    year = "2016",
    note = "[Addendum: Phys.Rev.D 94, 049904 (2016)]"
}

@article{Rodrigues:2018bdc,
    author = "Rodrigues, Manuel E. and de Sousa Silva, Marcos V.",
    title = "{Bardeen Regular Black Hole With an Electric Source}",
    eprint = "1802.05095",
    archivePrefix = "arXiv",
    primaryClass = "gr-qc",
    doi = "10.1088/1475-7516/2018/06/025",
    journal = "JCAP",
    volume = "06",
    pages = "025",
    year = "2018"
}

@article{Kumar:2020yem,
    author = "Kumar, Rahul and Kumar, Amit and Ghosh, Sushant G.",
    title = "{Testing Rotating Regular Metrics as Candidates for Astrophysical Black Holes}",
    eprint = "2006.09869",
    archivePrefix = "arXiv",
    primaryClass = "gr-qc",
    doi = "10.3847/1538-4357/ab8c4a",
    journal = "Astrophys. J.",
    volume = "896",
    number = "1",
    pages = "89",
    year = "2020"
}

@article{Cataldo:1999wr,
    author = "Cataldo, Mauricio and Garcia, Alberto",
    title = "{Three dimensional black hole coupled to the Born-Infeld electrodynamics}",
    eprint = "hep-th/9903257",
    archivePrefix = "arXiv",
    doi = "10.1016/S0370-2693(99)00441-4",
    journal = "Phys. Lett. B",
    volume = "456",
    pages = "28--33",
    year = "1999"
}

@article{Toshmatov:2018tyo,
    author = "Toshmatov, Bobir and Stuchl{\'\i}k, Zden{\v{e}}k and Schee, Jan and Ahmedov, Bobomurat",
    title = "{Electromagnetic perturbations of black holes in general relativity coupled to nonlinear electrodynamics}",
    eprint = "1805.00240",
    archivePrefix = "arXiv",
    primaryClass = "gr-qc",
    doi = "10.1103/PhysRevD.97.084058",
    journal = "Phys. Rev. D",
    volume = "97",
    number = "8",
    pages = "084058",
    year = "2018"
}

@article{Cataldo:2000ns,
    author = "Cataldo, Mauricio and Garcia, Alberto",
    title = "{Regular (2+1)-dimensional black holes within nonlinear electrodynamics}",
    eprint = "hep-th/0004177",
    archivePrefix = "arXiv",
    doi = "10.1103/PhysRevD.61.084003",
    journal = "Phys. Rev. D",
    volume = "61",
    pages = "084003",
    year = "2000"
}

@article{Novello:2000km,
    author = "Novello, M. and Perez Bergliaffa, Santiago E. and Salim, J. M.",
    title = "{Singularities in general relativity coupled to nonlinear electrodynamics}",
    eprint = "gr-qc/0003052",
    archivePrefix = "arXiv",
    doi = "10.1088/0264-9381/17/18/316",
    journal = "Class. Quant. Grav.",
    volume = "17",
    pages = "3821--3832",
    year = "2000"
}

@article{EventHorizonTelescope:2019ggy,
    author = "Akiyama, Kazunori and others",
    collaboration = "Event Horizon Telescope",
    title = "{First M87 Event Horizon Telescope Results. VI. The Shadow and Mass of the Central Black Hole}",
    eprint = "1906.11243",
    archivePrefix = "arXiv",
    primaryClass = "astro-ph.GA",
    doi = "10.3847/2041-8213/ab1141",
    journal = "Astrophys. J. Lett.",
    volume = "875",
    number = "1",
    pages = "L6",
    year = "2019"
}

@article{Gralla:2019xty,
    author = "Gralla, Samuel E. and Holz, Daniel E. and Wald, Robert M.",
    title = "{Black Hole Shadows, Photon Rings, and Lensing Rings}",
    eprint = "1906.00873",
    archivePrefix = "arXiv",
    primaryClass = "astro-ph.HE",
    doi = "10.1103/PhysRevD.100.024018",
    journal = "Phys. Rev. D",
    volume = "100",
    number = "2",
    pages = "024018",
    year = "2019"
}

@article{Cunha:2018acu,
    author = "Cunha, Pedro V. P. and Herdeiro, Carlos A. R.",
    title = "{Shadows and strong gravitational lensing: a brief review}",
    eprint = "1801.00860",
    archivePrefix = "arXiv",
    primaryClass = "gr-qc",
    doi = "10.1007/s10714-018-2361-9",
    journal = "Gen. Rel. Grav.",
    volume = "50",
    number = "4",
    pages = "42",
    year = "2018"
}

@article{Johannsen:2010ru,
    author = "Johannsen, Tim and Psaltis, Dimitrios",
    title = "{Testing the No-Hair Theorem with Observations in the Electromagnetic Spectrum: II. Black-Hole Images}",
    eprint = "1005.1931",
    archivePrefix = "arXiv",
    primaryClass = "astro-ph.HE",
    doi = "10.1088/0004-637X/718/1/446",
    journal = "Astrophys. J.",
    volume = "718",
    pages = "446--454",
    year = "2010"
}

@article{Grenzebach:2014fha,
    author = {Grenzebach, Arne and Perlick, Volker and L{\"a}mmerzahl, Claus},
    title = "{Photon Regions and Shadows of Kerr-Newman-NUT Black Holes with a Cosmological Constant}",
    eprint = "1403.5234",
    archivePrefix = "arXiv",
    primaryClass = "gr-qc",
    doi = "10.1103/PhysRevD.89.124004",
    journal = "Phys. Rev. D",
    volume = "89",
    number = "12",
    pages = "124004",
    year = "2014"
}

@article{Perlick:2015vta,
    author = "Perlick, Volker and Tsupko, Oleg Yu. and Bisnovatyi-Kogan, Gennady S.",
    title = "{Influence of a plasma on the shadow of a spherically symmetric black hole}",
    eprint = "1507.04217",
    archivePrefix = "arXiv",
    primaryClass = "gr-qc",
    doi = "10.1103/PhysRevD.92.104031",
    journal = "Phys. Rev. D",
    volume = "92",
    number = "10",
    pages = "104031",
    year = "2015"
}

@article{EventHorizonTelescope:2021dqv,
    author = "Kocherlakota, Prashant and others",
    collaboration = "Event Horizon Telescope",
    title = "{Constraints on black-hole charges with the 2017 EHT observations of M87*}",
    eprint = "2105.09343",
    archivePrefix = "arXiv",
    primaryClass = "gr-qc",
    reportNumber = "FERMILAB-PUB-21-847-PPD",
    doi = "10.1103/PhysRevD.103.104047",
    journal = "Phys. Rev. D",
    volume = "103",
    number = "10",
    pages = "104047",
    year = "2021"
}

@article{Younsi:2016azx,
    author = "Younsi, Ziri and Zhidenko, Alexander and Rezzolla, Luciano and Konoplya, Roman and Mizuno, Yosuke",
    title = "{New method for shadow calculations: Application to parametrized axisymmetric black holes}",
    eprint = "1607.05767",
    archivePrefix = "arXiv",
    primaryClass = "gr-qc",
    doi = "10.1103/PhysRevD.94.084025",
    journal = "Phys. Rev. D",
    volume = "94",
    number = "8",
    pages = "084025",
    year = "2016"
}

@article{Bambi:2015kza,
    author = "Bambi, Cosimo",
    title = "{Testing black hole candidates with electromagnetic radiation}",
    eprint = "1509.03884",
    archivePrefix = "arXiv",
    primaryClass = "gr-qc",
    doi = "10.1103/RevModPhys.89.025001",
    journal = "Rev. Mod. Phys.",
    volume = "89",
    number = "2",
    pages = "025001",
    year = "2017"
}

@article{Broderick:2013rlq,
    author = "Broderick, Avery E. and Johannsen, Tim and Loeb, Abraham and Psaltis, Dimitrios",
    title = "{Testing the No-Hair Theorem with Event Horizon Telescope Observations of Sagittarius A*}",
    eprint = "1311.5564",
    archivePrefix = "arXiv",
    primaryClass = "astro-ph.HE",
    doi = "10.1088/0004-637X/784/1/7",
    journal = "Astrophys. J.",
    volume = "784",
    pages = "7",
    year = "2014"
}

@article{Johannsen:2015hib,
    author = "Johannsen, Tim and Broderick, Avery E. and Plewa, Philipp M. and Chatzopoulos, Sotiris and Doeleman, Sheperd S. and Eisenhauer, Frank and Fish, Vincent L. and Genzel, Reinhard and Gerhard, Ortwin and Johnson, Michael D.",
    title = "{Testing General Relativity with the Shadow Size of Sgr A*}",
    eprint = "1512.02640",
    archivePrefix = "arXiv",
    primaryClass = "astro-ph.GA",
    doi = "10.1103/PhysRevLett.116.031101",
    journal = "Phys. Rev. Lett.",
    volume = "116",
    number = "3",
    pages = "031101",
    year = "2016"
}

@article{Vagnozzi:2020quf,
    author = "Vagnozzi, Sunny and Bambi, Cosimo and Visinelli, Luca",
    title = "{Concerns regarding the use of black hole shadows as standard rulers}",
    eprint = "2001.02986",
    archivePrefix = "arXiv",
    primaryClass = "gr-qc",
    doi = "10.1088/1361-6382/ab7965",
    journal = "Class. Quant. Grav.",
    volume = "37",
    number = "8",
    pages = "087001",
    year = "2020"
}

@article{Chen:2022nbb,
    author = "Chen, Yifan and Roy, Rittick and Vagnozzi, Sunny and Visinelli, Luca",
    title = "{Superradiant evolution of the shadow and photon ring of Sgr A{\ensuremath{\star}}}",
    eprint = "2205.06238",
    archivePrefix = "arXiv",
    primaryClass = "astro-ph.HE",
    doi = "10.1103/PhysRevD.106.043021",
    journal = "Phys. Rev. D",
    volume = "106",
    number = "4",
    pages = "043021",
    year = "2022"
}

@article{Hennigar:2018hza,
    author = "Hennigar, Robie A. and Poshteh, Mohammad Bagher Jahani and Mann, Robert B.",
    title = "{Shadows, Signals, and Stability in Einsteinian Cubic Gravity}",
    eprint = "1801.03223",
    archivePrefix = "arXiv",
    primaryClass = "gr-qc",
    doi = "10.1103/PhysRevD.97.064041",
    journal = "Phys. Rev. D",
    volume = "97",
    number = "6",
    pages = "064041",
    year = "2018"
}

@article{Kuang:2022ojj,
    author = "Kuang, Xiao-Mei and Tang, Zi-Yu and Wang, Bin and Wang, Anzhong",
    title = "{Constraining a modified gravity theory in strong gravitational lensing and black hole shadow observations}",
    eprint = "2206.05878",
    archivePrefix = "arXiv",
    primaryClass = "gr-qc",
    doi = "10.1103/PhysRevD.106.064012",
    journal = "Phys. Rev. D",
    volume = "106",
    number = "6",
    pages = "064012",
    year = "2022"
}

@article{Bronzwaer:2021lzo,
    author = "Bronzwaer, Thomas and Falcke, Heino",
    title = "{The Nature of Black Hole Shadows}",
    eprint = "2108.03966",
    archivePrefix = "arXiv",
    primaryClass = "astro-ph.HE",
    doi = "10.3847/1538-4357/ac1738",
    journal = "Astrophys. J.",
    volume = "920",
    number = "2",
    pages = "155",
    year = "2021"
}

\end{document}